\documentclass[letterpaper,twocolumn,10pt]{article}
\usepackage{usenix}
\usepackage{tikz}
\usepackage{amsmath}

\usepackage{filecontents}
\usepackage{tabularx, multirow, amsmath, pifont}
\usepackage{makecell}
\usepackage{listings}
\usepackage{arydshln}
\usepackage{amssymb}
\usepackage{soul}
\usepackage{xcolor}
\usepackage{longtable,booktabs}
\usepackage{xtab}
\usepackage[most]{tcolorbox}
\usepackage[ruled,vlined, linesnumbered]{algorithm2e}
\usepackage{caption}
\usepackage{algorithmic}
\usepackage{pgfplots}
\usepackage{pgfplotstable}
\usepackage{overpic}
\pgfplotsset{compat=1.17}
\newcommand{\xmark}{\ding{55}}%

\definecolor{codegreen}{rgb}{0,0.6,0}
\definecolor{codegray}{rgb}{0.5,0.5,0.5}
\definecolor{codepurple}{rgb}{0.58,0,0.82}
\definecolor{backcolour}{rgb}{0.95,0.95,0.92}

\lstdefinestyle{mystyle}{
commentstyle=\color{codegreen},
  keywordstyle=\color{magenta},
  numberstyle=\tiny\color{codegray},
  stringstyle=\color{codepurple},
  basicstyle=\ttfamily\footnotesize,
  breakatwhitespace=false,         
  breaklines=true,                 
  captionpos=b,                    
  keepspaces=true,                 
  numbers=left, 
  numbersep=5pt,                  
  showspaces=false,                
  showstringspaces=false,
  showtabs=false,                  
  tabsize=2
}

\def\framework{RAGent}
\def\gitrepo{\url{https://github.com/accessframework/RAGent}}

\begin{document}

\date{}

\title{\Large \bf RAGent: Retrieval-based Access Control Policy Generation}


\author{
{\rm Sakuna Harinda Jayasundara}\\
The School of Computer Science\\
University of Auckland
\and
{\rm Nalin Asanka Gamagedara Archchilage}\\
The School of Computing Technologies\\
RMIT University
\and
{\rm Giovanni Russello}\\
The School of Computer Science\\
University of Auckland
} 

\maketitle
\begin{abstract}
Manually generating access control policies from an organization’s high-level requirement specifications poses significant challenges. 
It requires laborious efforts to sift through multiple documents containing such specifications and translate their access requirements into access control policies. 
Also, the complexities and ambiguities of these specifications often result in errors by system administrators during the translation process, leading to data breaches. 
However, the automated policy generation frameworks designed to help administrators in this process are unreliable due to limitations, such as the lack of domain adaptation.
Therefore, to improve the reliability of access control policy generation, we propose \framework, a novel retrieval-based access control policy generation framework based on language models. 
\framework{} identifies access requirements from high-level requirement specifications with an average state-of-the-art F1 score of 87.9\%.
Through retrieval augmented generation, \framework{} then translates the identified access requirements into access control policies with an F1 score of 77.9\%.
Unlike existing frameworks, \framework{} generates policies with complex components like purposes and conditions, in addition to subjects, actions, and resources.
Moreover, \framework{} automatically verifies the generated policies and iteratively refines them through a novel verification-refinement mechanism, further improving the reliability of the process by $\approx 3\%$, reaching the F1 score of 80.6\%.
We also introduce three annotated datasets for developing access control policy generation frameworks in the future, addressing the data scarcity of the domain.
\end{abstract}

\section{Introduction}
\label{sec:introduction}


According to the 2023 Verizon data breach incident report, 74\% of data breaches involve human elements \cite{verizon}.
These include accidental human errors system administrators make that can result in severe data breaches \cite{Page_2023}.
For example, in July 2023, Microsoft AI researchers released a public GitHub repository, providing users access to several AI models via an Azure storage bucket URL (Uniform Resource Locator) \cite{Page_2023}.
However, instead of allowing users only to download the models, the storage account administrator had mistakenly given ``full access'' to the entire Azure storage account \cite{Page_2023}.
Consequently, 38TB of company data, including passwords for Microsoft services, were leaked to the general public \cite{Page_2023}.

Such incidents based on human errors often occur when system administrators manually generate access control policies from the organization's high-level requirement specifications \cite{jayasundaravision, horstmann2024those, bauer2009real}.
These requirements are written by security experts \cite{bauer2009real, horstmann2024those} in documents explaining how information access should be managed within the organization \cite{narouei2018automatic}. 
However, they are often unstructured, ambiguous, and written in legal language that non-security experts like system administrators find difficult to understand \cite{horstmann2024those, narouei2018automatic}. 
Consequently, manually identifying access control requirements from these documents and generating access control policies become laborious and error-prone, resulting in human errors, that may lead to data breaches \cite{jayasundara2023sok, jayasundaravision, narouei2018automatic}.

Therefore, to help administrators correctly generate access control policies, previous research has introduced automated policy generation frameworks \cite{xia2022automated, narouei2015automatic, narouei2015towards, narouei2017identification, heaps2021access, alohaly2019automated, alohaly2018deep, alohaly2019towards, narouei2018automatic, slankas2014relation, slankas2013access, slankas2013accessid, slankas2012classifying}. 
Those frameworks automatically process high-level requirement specification documents, identify Natural Language Access Control Policies (NLACPs), and extract useful information (e.g., policy components like subjects, actions, and resources) for building access control policies from them, using machine learning (ML) and natural language processing (NLP) techniques \cite{slankas2014relation, xiao2012automated, narouei2015automatic, slankas2013accessid, xia2022automated}.
They remove the human factor almost entirely from the policy generation process, attempting to reduce access control failures due to human mistakes \cite{kaur2021human}.


However, the existing policy generation frameworks are far from accurate due to their limitations, indicating that they cannot be used without human involvement \cite{del2022systematic, jayasundara2023sok, narouei2015automatic, jayasundaravision, narouei2018automatic}.
For instance, most deep learning-based frameworks were not specifically adapted for extracting access control policy components from NLACPs \cite{narouei2015automatic, xia2022automated, alohaly2019automated, alohaly2018deep}. 
Instead, they mostly rely on general-purpose NLP models like Semantic Role Labeling (SRL) models \cite{xia2022automated, narouei2015automatic, narouei2018automatic}, which are not tailored for extracting policy components from NLACPs.
As a result, they often struggle to accurately extract policy components and sometimes extract incorrect or unrelated entities from NLACPs, resulting in low precision and recall at the end \cite{narouei2015automatic, xia2022automated}.
Furthermore, these frameworks cannot extract intricate policy components like purposes and conditions from NLACPs \cite{ xia2022automated, heaps2021access, slankas2014relation, narouei2015automatic, slankas2013accessid, slankas2013access, narouei2017towards, narouei2018automatic, xiao2012automated}. 
Therefore, they may generate incomplete access control policies that overlook crucial purposes and conditions necessary for adherence to high-level organizational requirements \cite{alohaly2019towards}.

Out of all such limitations, the most crucial limitation of the existing frameworks is their inability to adapt to organization-specific information, such as pre-defined users, resources, etc., when generating policies from NLACPs, without re-training the entire framework \cite{xia2022automated, heaps2021access}.
As a result, when used in an organization whose confidential access control information is not part of the framework's training data, the existing frameworks often generate policies with entities that do not match the organization's authorization system. 
Finally, none of the existing access control policy generation frameworks verify the automatically generated policies and refine any incorrectly generated policies before adding them to the authorization system.
Consequently, it reduces the reliability of the policy generation process due to the incorrect policies generated by frameworks, increasing the chance of access control failures \cite{jayasundaravision}.
These limitations of existing access control policy generation frameworks collectively contribute to their low reliability, making them unsuitable for a reliable access control policy generation \cite{narouei2015automatic, del2022systematic, xia2022automated}.

Therefore, to improve the access control policy generation reliability, we propose \framework, a novel access control policy generation framework, making the following contributions.
\begin{itemize}
    \item We introduce \framework{}, a novel \ul{R}etrieval-based \ul{A}ccess control policy \ul{Gen}era\ul{t}ion framework to identify and translate NLACPs into access control policies using transformer-based LMs.
    Utilizing ``small'' open-source LMs compared to proprietary LMs, \framework{} enables its efficient and local deployment within the organization, ensuring the confidentiality of the organizational access control requirements \cite{Dagdelen_Dunn_Lee_Walker_Rosen_Ceder_Persson_Jain_2024} (Section \ref{sec:approach}). 
    \item We show that incorporating organization-specific information to generate access control policies through Retrieval Augmented Generation (RAG) \cite{lewis2020retrieval} helps improve the reliability of the access control policy generation process (Section \ref{sec:approach} and Appendix \ref{subsec:ablation/retrieval}).
    Furthermore, it enables the reliable adaptation of \framework{} to any new domains without costly training of ML models.
    \item We introduce a novel policy verification-refinement mechanism, improving the reliability of the process even further. It automatically verifies the generated policy and refines it iteratively using feedback, if it is generated incorrectly (Section \ref{sec:approach} and Appendix \ref{subsec:ablation/refinement}).
    \item We extensively evaluate \framework{} through real-world high-level requirement specifications \cite{slankas2014relation, xiao2012automated}, demonstrating that it identifies NLACPs with a state-of-the-art average F1 score of 87.9\%.
    Furthermore, in contrast to existing frameworks, \framework{} translates NLACPs into granular access control policies extracting five policy components (i.e., subjects, actions, resources, purpose, and conditions) with an average state-of-the-art F1 score of 80.7\%, which is 39.1\% higher than the current state-of-the-art \cite{xia2022automated} (Section \ref{sec:eval_results}).
    \item By applying \framework{} to a real-world policy generation scenario, we further highlight its ability to deal with complex access requirements involving multiple and hidden/implicit Access Control Rules (ACRs), which cannot be achieved using existing frameworks (Appendix \ref{sec: usecase}).
    \item We develop and release three annotated datasets by addressing the data scarcity in the access control policy generation domain (Section \ref{sec:datasets}).
    \item We make the implementation of \framework{} public via \gitrepo. 
\end{itemize}
\section{Related Work}
\label{sec:related_works}
\newcolumntype{M}[1]{>{\centering\arraybackslash}m{#1}}

\begin{table*}
  \centering
  \caption{Comparison of \framework{} to the existing automated policy generation frameworks. S - Subject, A - Action, R - Resource, P - Purpose, C - Condition}
    \label{tab:related_works}
  \begin{tabular}{lcccccccccc}
    \toprule
    \multirow{2}{0.5cm}{} &
    \multirow{2}{1.7cm}{\centering \textbf{Domain Adaptation}} &
    \multirow{2}{2cm}{\centering \textbf{NLACP identification}} &
      \multirow{2}{2.2cm}{\centering \textbf{ACR decision extraction}} &
      \multicolumn{5}{M{2.6cm}}{\textbf{Access control policy generation}} &
      \multirow{2}{3cm}{\centering \textbf{Policy verification and refinement}} \\
      \vspace{3pt}
    & & & & S & A & R & P & C & \\
    \midrule
    Text2Policy \cite{xiao2012automated} & \xmark & \checkmark & \xmark & \checkmark & \checkmark & \checkmark & \xmark & \xmark & \xmark \\
    ACRE \cite{slankas2013access, slankas2013accessid, slankas2014relation} & \xmark &\checkmark & \xmark & \checkmark & \checkmark & \checkmark & \xmark & \xmark & \xmark \\
    Narouei et. al. \cite{narouei2015automatic, narouei2015towards, narouei2017towards, narouei2018automatic} & \xmark & \checkmark & \xmark & \checkmark & \checkmark & \checkmark & \xmark & \xmark & \xmark \\
    Xia et. al. \cite{xia2022automated} & \xmark & \checkmark & \xmark & \checkmark & \checkmark & \checkmark & \xmark & \xmark & \xmark \\
    \midrule
    \textbf{\framework} & \textbf{\checkmark} & \textbf{\checkmark} & \textbf{\checkmark} & \textbf{\checkmark} & \textbf{\checkmark} & \textbf{\checkmark} & \textbf{\checkmark} & \textbf{\checkmark} & \textbf{\checkmark} \\
    
    \bottomrule
  \end{tabular}
\end{table*}

Previous research has proposed numerous automated policy generation frameworks \cite{xiao2012automated, slankas2012classifying, slankas2013access, slankas2013accessid, slankas2014relation, narouei2015automatic, narouei2015towards, narouei2018automatic, narouei2017identification, narouei2017towards, xia2022automated, heaps2021access}, aiming to identify NLACPs from high-level requirement specification documents and extract their policy components.
They often follow a three-step approach: (1) pre-processing \cite{slankas2012classifying, slankas2014relation, slankas2013access}, (2) NLACP identification using text classification techniques such as Support Vector Machines (SVM) \cite{slankas2013access, slankas2014relation}, and (3) policy components (i.e., subjects, actions, and resources) extraction through information extraction techniques such as SRL \cite{ xia2022automated, narouei2017towards, narouei2015automatic}.
System administrators can then use the extracted policy components to formulate machine-executable policies in standard access control languages such as XACML (eXtensible Access Control Markup Language) \cite{xacml}. 

Xiao et al. developed Text2Policy \cite{xiao2012automated}, utilizing syntactic pattern matching to identify NLACPs and shallow parsing based on four semantic patterns to extract policy components from them. 
However, its reliance on a limited set of semantic patterns restricts its ability to extract components from NLACPs that do not conform to these patterns.

Therefore, Slankas et al. \cite{slankas2012classifying, slankas2013access, slankas2013accessid, slankas2014relation} proposed a policy generation framework, Access Control Relation Extraction (ACRE) as a more reliable alternative to Text2Policy. 
ACRE utilizes the k-Nearest Neighbours (k-NN) algorithm to identify NLACPs with an average F1 score of 72\%.
Moreover, it employs dependency parsing with bootstrapping \cite{de2006generating} to extract ACRs and their policy components from NLACPs based on their grammatical structures \cite{slankas2014relation}.
The bootstrapping technique enables ACRE to dynamically expand the dependency pattern database (i.e., grammatical relationships used for parsing NLACPs and extracting ACRs) according to the given high-level requirement specification document. 
Therefore, ACRE would not be limited to a pre-defined set of patterns as seen in Text2Policy \cite{slankas2014relation}.
As a result, Slankas et al. achieved an average F1-score of 57.2\% for policy component extraction \cite{slankas2014relation}.
However, ACRE \cite{slankas2012classifying, slankas2013access, slankas2013accessid, slankas2014relation} needs repetition of sentence structures throughout the high-level requirement specification document to perform optimally \cite{narouei2018automatic}.
Additionally, since ACRE relies on dependency parsing, it may struggle to accurately identify ``subjects'' and ``resources'' as nouns and ``actions'' as verbs in complex and ambiguous NLACPs \cite{slankas2014relation, nivre2008algorithms}.
As a result, incorrect ACRs may be extracted, leading to low reliability of policy generation \cite{jayasundara2023sok}.

Therefore, by attempting to improve the reliability of policy generation without relying on pre-defined grammar, researchers then utilized deep-learning techniques to develop access control policy generation frameworks \cite{narouei2015automatic, narouei2015towards, narouei2018automatic, narouei2017identification, narouei2017towards, alohaly2018deep, alohaly2019automated, alohaly2019towards, xia2022automated, heaps2021access}. 
For instance, Narouei et al. used deep learning-based SRL to extract ACRs and their policy components \cite{narouei2015automatic, narouei2015towards, narouei2017towards, narouei2018automatic}.
They used SENNA (Semantic/syntactic Extraction using a Neural Network Architecture) \cite{collobert2011natural} SRL model to identify the predicate-argument structure of NLACPs and extract subject (ARG0) and object (ARG1) for a given predicate \cite{narouei2018automatic}.
Using SENNA, Narouei et al. achieved an overall policy component extraction F1 score of 69.8\% \cite{narouei2018automatic}, which is an improvement of 12.6\% compared to ACRE \cite{slankas2014relation}. 
Similarly, Xia et al. \cite{xia2022automated} utilized a BERT-based SRL model \cite{shi2019simple, zhang2020semantics} to extract ACRs, attaining an average F1 score of 72\% on the same dataset. 

Nevertheless, existing deep learning-based frameworks lack domain adaptation due to the scarcity of domain-related datasets \cite{narouei2015towards, xia2022automated, narouei2017towards}.
Therefore, they sometimes fail to accurately extract access control policy components from each ACR of NLACPs, 
leading to low reliability (e.g., the highest policy component extraction F1 score so far is 72\% \cite{xia2022automated}) \cite{xia2022automated, narouei2015automatic}.
For example, generic deep learning-based SRL models used to extract policy components mainly identify one subject and one object for each predicate of the given sentence \cite{shi2019simple}. 
This predicate may not always pertain to an ``action'' related to access control (e.g., can, could, etc.) \cite{xia2022automated}. 
As a result, SRL models extract subjects and resources related to such incorrect predicates, lowering the reliability of ACR extraction \cite{narouei2018automatic}.
Even if these frameworks are adapted to the access control domain, once they are deployed in a new organization, they might still not generate accurate policies.
This is because they are not aware of the organization-specific information, such as users/roles and resources defined in the organization, that directly affect policy generation.
As a result, the policy generation frameworks may generate policies by extracting entities from the NLACPs solely based on their grammatical properties that may not match the organization's authorization system.
For instance, for the NLACP, \emph{``The senior lecturer can read student grades.}, the policy generation frameworks might generate \emph{``lecturer''} as the subject, even though the pre-defined entity in the system is \emph{``senior lecturer''}.
While such situations can be avoided by training these frameworks using the organization's access requirements, it can be costly \cite{firstpost} and should be done repetitively each time new roles and resources are introduced to the system.

Moreover, the existing access control policy generation frameworks can only extract three main policy components, subject, action, and resource, from NLACPs. 
These frameworks fall short when it comes to complex NLACPs containing ACRs with different access decisions and complex policy components, like conditions and purposes \cite{narouei2015automatic, narouei2015towards, narouei2018automatic, narouei2017identification, narouei2017towards, alohaly2018deep, alohaly2019automated, alohaly2019towards, xia2022automated, heaps2021access, slankas2014relation, slankas2013access, xiao2012automated}.
For example, consider the NLACP \emph{``The nurse cannot read the records to prescribe medicine, but the doctor can''}.
None of the existing frameworks can detect the purpose of the NLACP as \emph{``prescribe medicine''}, generating policies that allow access requests without checking their purposes and conditions.
This may result in insider attacks due to privilege misuse \cite{bauer2009real}.
Also, the existing frameworks \cite{xia2022automated, xiao2012automated} provide an overall access decision for the NLACP, irrespective of the individual access decisions of the ACRs belong to the NLACP (e.g., \emph{``The [\ul{nurse cannot read the records to prescribe medicine}]$_{DENY}$, but the [\ul{doctor can}]$_{ALLOW}$''}).
Consequently, if the NLACP involves different access decisions, existing frameworks may always generate incorrect ACRs, causing access control failures. 


All the above limitations highlight that the complete automation of the access control policy generation using existing frameworks is far from reliable.
Therefore, for a reliable access control policy generation, \emph{incorrectly generated policies} by automated frameworks should be refined automatically or manually by the system administrator before adding them to the authorization system \cite{jayasundara2023sok, jayasundaravision}.
To this end, policy generation frameworks must identify whether or not the generated policy is incorrect and the reasons for being incorrect and provide feedback \cite{xu2017system}. 
However, none of the existing frameworks have performed these crucial steps for handling incorrectly generated policies \cite{xiao2012automated, slankas2014relation, narouei2015automatic, narouei2018automatic, alohaly2019automated, xia2022automated, heaps2021access}. 
Therefore, this remains yet another significant gap that needs addressing to enhance the reliability of access control policy generation.


 This paper addresses those research gaps by introducing \framework, a novel retrieval-based access control policy generation framework, leveraging transformer-based LMs and their impressive language understanding capabilities \cite{vaswani2017attention}. 
As shown in Table \ref{tab:related_works}, \framework{} stands out from existing policy generation frameworks in several key ways.
Firstly, \framework{} accurately generates access control policies from high-level requirement specifications. 
It adeptly manages complex access requirements, including those with intricate components like purposes and conditions, and also with ACRs containing different access decisions.
Furthermore, \framework{} performs RAG incorporating organization-specific information like users and resources in policy generation to produce reliable policies that match the organization's authorization system (i.e., domain adaptation).
Additionally, it goes a step further by automatically verifying the generated policies and offering feedback in case of errors. 
This feedback is first used to iteratively refine the generated policy automatically and also manually by the system administrator if the automatic refinement fails.



\section{\framework}
\label{sec:approach}
\begin{figure*}[h]
    \centering
    \includegraphics[height=11.8cm]{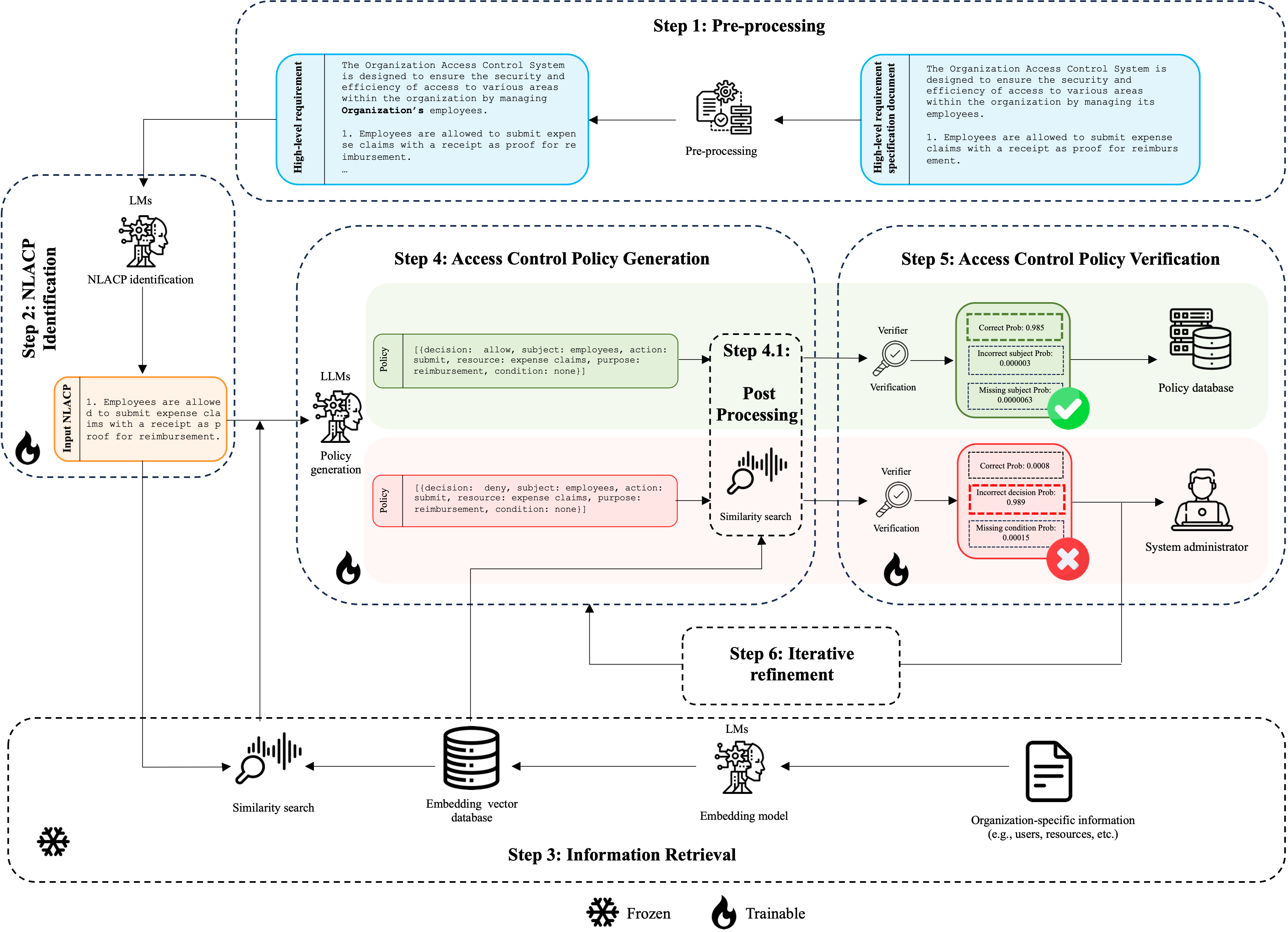}
    
    \caption{High-level architecture of \framework{}. \textbf{Step 1}: Pre-processing. \textbf{Step 2}: NLACP identification. \textbf{Step 3}: Retrieving information relevant to generating the access control policy of the NLACP. \textbf{Step 4}: Generating access control policy using the retrieved information. \textbf{Step 4.1}: Post-processing the generated policy. \textbf{Step 5}: Access control policy verification that allows the correctly generated policies to apply to the system and provides feedback if the generated policy is found incorrect. \textbf{Step 6}: Iteratively refine the generated policy using the verification feedback.
    }
    
    \label{fig:framework}
\end{figure*}

The main objective of \framework, is to improve the reliability of the access control policy generation process.
It utilizes transformer-based LMs to generate access control policies in six steps, as shown in Figure \ref{fig:framework}.
\begin{enumerate}
    \item Step 1: Pre-processing
    \item Step 2: NLACP Identification
    \item Step 3: Information Retrieval
    \item Step 4: Access Control Policy Generation
    \begin{enumerate}
        \item Step 4.1 Post-processing
    \end{enumerate}
    \item Step 5: Access Control Policy Verification
    \item Step 6: Iterative Refinement
\end{enumerate}

According to Figure \ref{fig:framework}, the input to \framework{} is the organization's high-level requirement specification document (i.e., input document) written by a security expert in NL. 
In this research, we assume that the input document does not contain any requirement conflicts, similar to previous research \cite{narouei2015automatic, xia2022automated, narouei2018automatic, slankas2013access}. 
First, \framework{} pre-processes the sentences of the input document.
Then, \framework{} classifies the pre-processed sentences to identify NLACPs in the second step. 
Once an NLACP is identified, in the third step, \framework{} retrieves organization-specific information relevant to translating the identified NLACP into an access control policy.
This information will be combined with the NLACP and fed to the policy generation module in the fourth step to generate its access control policy containing ACRs with five policy components (i.e., subjects, actions, resources, purposes, and conditions) and their rule decisions (i.e., allow or deny) through RAG.
\framework{} generates the access control policy as a structured representation of the NLACP \cite{Dagdelen_Dunn_Lee_Walker_Rosen_Ceder_Persson_Jain_2024}. 
By translating the NLACP into a structured representation rather than utilizing a standard access control language, \framework{} becomes universally applicable.
It enables any organization, regardless of the language utilized in their authorization system, to use \framework, and seamlessly transform the generated representation into their specific access control language.
Then, \framework{} post-processes the generated policy to ensure that it only contains policy components that align with the authorization system.

After generating the access control policy, it is verified in the fifth step using a novel policy verification technique to decide whether or not it is correct.
If the generated access control policy is verified as correct, it can be applied to the authorization system.
On the other hand, if the policy is verified as incorrect, the verifier provides the reason (i.e., error type) for it being incorrect as feedback.
In that case, as shown in Figure \ref{fig:framework}, the feedback is used to address the identified error(s) iteratively and generate the correct refined policy within $n$ iterations.
If the policy is still incorrect even after $n$ rounds, it will be sent to the administrator with feedback 
for the manual refinement. 
This helps administrators easily find erroneous ACRs and refine them before adding them to the authorization system.

\subsection{Step 1: Pre-processing}
\label{subsec:approach/preprocessing}

Once a high-level requirement specification document is provided as the input, \framework{} first pre-processes 
it by segmenting its paragraphs. 
Then, \framework{} performs coreference resolution on those paragraphs, which has rarely been performed in existing policy generation frameworks \cite{narouei2015automatic, slankas2014relation}.

Coreference resolution is the task of determining whether or not the two expressions refer to the same entity in a sentence or a paragraph \cite{narouei2018automatic}.
For example, consider an NLACP, \emph{``Nurses are allowed to read the prescription, but they are not allowed to change it."}. 
In the above NLACP, \emph{``they"} refers to \emph{``Nurses"} and \emph{``it"} refers to \emph{``the prescription"}. 
However, without the first rule (i.e., \emph{``Nurses are allowed to read the prescription"}), the second rule loses its meaning (i.e., \emph{``They are not allowed to change it"}) since \emph{``They"} and \emph{``it"} are unknown. 
Therefore, once the coreferences are resolved, the NLACP will be improved into \emph{``Nurses are allowed to read the prescription, but \underline{Nurses} are not allowed to change \underline{the prescription}."}, allowing the framework to extract more meaningful and exclusive ACRs and generate policies.

After the coreference resolution, \framework{} further segment the paragraphs into sentences and tokenize them to be classified in Step 2 to identify NLACPs. 


\subsection{Step 2: NLACP Identification}
\label{subsec:approach/identification}

High-level requirement specification documents contain access control requirements (i.e., NLACP) as well as sentences that are not related to access control (i.e., non-NLACP) \cite{narouei2018automatic}.
Therefore, only NLACPs should be identified 
to translate them into access control policies \cite{xiao2012automated, narouei2015automatic, narouei2018automatic, xia2022automated, heaps2021access}. 

We formulate this NLACP identification as a binary text classification task where the goal is to classify each sentence into either NLACP or non-NLACP categories \cite{xia2022automated, slankas2012classifying, slankas2013access, slankas2013accessid, slankas2014relation, narouei2015automatic, narouei2018automatic, heaps2021access}.
When classifying a sentence, it is important to incorporate the context from the entire sentence to decide its category \cite{gasparetto2022survey}.
However, traditional ML-based classification techniques such as SVM, k-NN, and Naïve Bayes used in previous research \cite{slankas2013access, slankas2013accessid, xiao2012automated, narouei2015automatic, narouei2015towards, narouei2017identification, narouei2018automatic} do not utilize the context from the entire sentence, missing dependencies between words of the sentence relevant to classify it correctly \cite{gasparetto2022survey}.
As a solution, in recent years, bi-directional transformer-based language models like BERT \cite{devlin2018bert} and RoBERTa \cite{liu2019roberta} have emerged as superior alternatives for text classification.
They rely on the ``self-attention'' mechanism \cite{vaswani2017attention} to learn dependencies between tokens/words of the sentence (i.e., learning context), exhibiting state-of-the-art performance in NLP tasks like sentiment analysis.
Therefore, in \framework{}, we use BERT LM to identify NLACPs from NL sentences  \cite{xia2022automated, heaps2021access}. 

However, we cannot use BERT directly to classify sentences as NLACPs, as it is pre-trained to predict the masked token of the input sentence (i.e., Masked Language Modeling) and to predict the next sentence (i.e., Next Sentence Prediction) \cite{devlin2018bert}.
Therefore, we improve the BERT LM by adding a feed-forward network (FFN) containing a single linear layer on top of the final hidden state of BERT corresponding to the $[CLS]$ token that provides an aggregate representation of the input sequence \cite{devlin2018bert}.
Then, we fine-tune BERT with the access control policy dataset (Section \ref{subsec:dataset/access_dataset}) containing NLACPs and non-NLACPs, minimizing the cross-entropy loss \cite{devlin2018bert}.
It allows BERT to learn how to identify NLACPs from non-NLACPs accurately \cite{xia2022automated, heaps2021access}.

\subsection{Step 3: Information Retrieval}
\label{subsec:approach/retrieval}

Even though the NLACPs are correctly identified in the previous step, they should be translated into access control policies accurately for them to be enforceable.
Also, the generated access control policies should address the organization's pre-defined entities (e.g., users and resources).
However, existing frameworks often fail to produce such access control policies when not trained on organization-specific access requirements, rendering them unenforceable \cite{narouei2018automatic}.
Also, that training should repeat each time the framework is being employed in new organizations (i.e., change in contexts) and each time new users/roles or resources are introduced to the organization's authorization system, which can be expensive.
Therefore, instead of training \framework{} (i.e., updating its parametric memory) repetitively to adapt it to specific contexts, \framework{} uses organization-specific information as a non-parametric memory \cite{ding2024survey}, to generate policies based on them (i.e., RAG).
To this end, we consider pre-defined subjects, actions, resources, purposes, and conditions as organization-specific information since they are the main policy components of an ACR \cite{brodie2005usable, brodie2006empirical}.

However, all the information, like all subjects and resources defined in the system, may not be relevant to generating an access control policy from a specific NLACP.
Therefore, to incorporate only required information to generate policies, \framework{} uses dense retrieval \cite{ding2024survey}.
Dense retrieval involves representing the organization-specific information as high-dimensional embedding vectors and retrieving them based on their similarity to the given NLACP \cite{ding2024survey}.
Since it considers the semantic meaning (encoded as embedding vectors) rather than word matches (as in sparse retrieval \cite{ding2024survey}) when retrieving information, it is also robust to lexical variations such as synonyms.
For example, when retrieving relevant subjects for translating the NLACP, \emph{``Graduate teaching assistant can read grades.''}, sparse retrieval techniques like BM25 \cite{robertson2009probabilistic} will search for the subjects containing words \emph{``graduate''}, \emph{``teaching''} or \emph{``assistant''}, overlooking synonymous subjects like \emph{``GTA''}.
In contrast, since dense retrieval uses semantic similarity between entities and NLACP via embeddings, it can identify such synonymous subjects even though they do not match the exact terms of the NLACP.

Therefore, to facilitate dense retrieval, first, \framework{} generates embedding vectors for subjects, actions, resources, purposes, and conditions of the organization. 
To this end, it utilizes an open-source lightweight embedding model \cite{emb2024mxbai} that shows superior performance in retrieval according to the Huggingface MTEB leaderboard \cite{muennighoff2022mteb} at the time of writing.
The embedding vectors are then stored in separate specialized databases (one for each information type), called ``vector databases'' \cite{han2023comprehensive} as shown in Figure \ref{fig:framework}. 
It helps optimize the retrieval of the stored information through algorithms like Product Quantization \cite{pinecone}. 
After preparing the vector databases, \framework{} generates the embedding vector for the identified NLACP using the same embedding model as shown in Figure \ref{fig:framework}. 
Then, it compares the NLACP's embedding vector with embedding vectors of the information from vector databases using cosine similarity \cite{gao2023retrieval, muennighoff2022mteb} to retrieve the most similar $k$ entities from each database \cite{ding2024survey}. 
These $5k$ entities are used in Step 4 to help generate the access control policy for an NLACP.

\subsection{Step 4: Access Control Policy Generation}
\label{subsec:approach/generation}

\begin{figure}[!ht]
    \centering
    \includegraphics[width=0.47\textwidth]{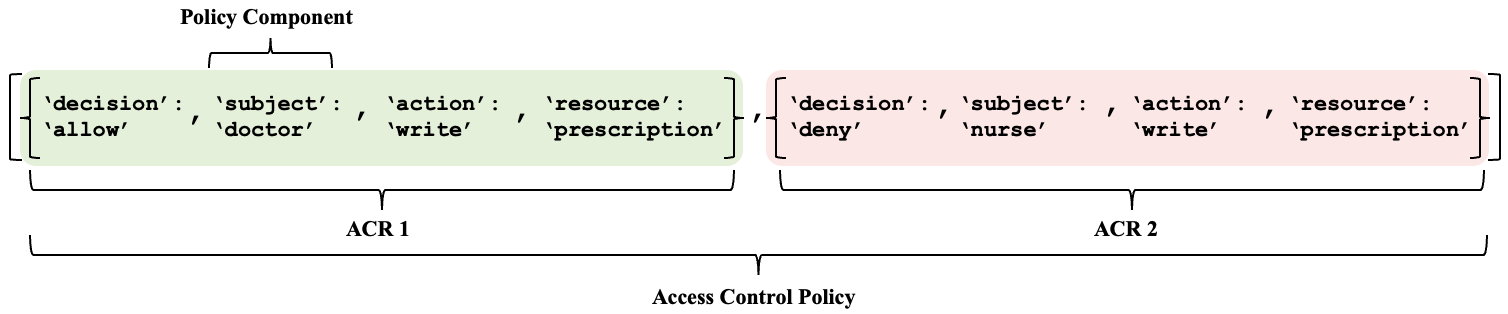}
    
    \caption{Structured representation of the NLACP \emph{``The doctor can write prescriptions, but the nurse cannot.''}.}
    
    \label{fig:template}
\end{figure}

After identifying the NLACPs of the high-level requirement specification documents (Step 2) with the relevant information (Step 3), \framework{} translates them into access control policies through \emph{structured information extraction} \cite{Dagdelen_Dunn_Lee_Walker_Rosen_Ceder_Persson_Jain_2024}, as shown in Figure \ref{fig:framework}.
In other words, \framework{} extracts information from an NLACP according to a structure/hierarchy (e.g., JSON) delineating an access control policy.
It represents underlying ACRs of the NLACP, access decisions, and policy components (i.e., subject, actions, resource, purpose, and condition) of each rule \cite{slankas2014relation, narouei2017towards}.
For example, consider an NLACP, \emph{``The doctor can write prescriptions, but the nurse cannot.''}. 
Given that NLACP, the \framework{} generates the access control policy according to a structure as shown in Figure \ref{fig:template}, maintaining the relationships between policy components within each ACR \cite{Dagdelen_Dunn_Lee_Walker_Rosen_Ceder_Persson_Jain_2024} (We removed the empty purpose and condition fields of each rule for clarity).
This can easily be transformed into any standard access control language, such as XACML \cite{brodie2006empirical}, when applying it to the authorization system.

To generate access control policies from NLACPs, previous research has often used general-purpose SRL models, leading to several limitations like extracting unwanted entities \cite{narouei2015automatic, narouei2018automatic} and not being able to extract purposes and conditions (Section \ref{sec:related_works}).  
Therefore, to avoid such limitations, in \framework, we utilize domain-adapted Large LMs (LLMs), which have been proven effective in structured information extraction from input texts \cite{Dagdelen_Dunn_Lee_Walker_Rosen_Ceder_Persson_Jain_2024}.  
They show high efficacy in tasks that require a combination of a natural and domain-specific language, as well as an understanding of specialized terminology after adapting them to a specific domain \cite{roziere2023code, narouei2018automatic}. 
As a result, domain-adapted LLMs have been successfully used in similar tasks to access control policy generation, such as code generation from NL descriptions \cite{roziere2023code, ni2023lever, zan2023large, llama3} and structured information extraction from scientific documents \cite{Dagdelen_Dunn_Lee_Walker_Rosen_Ceder_Persson_Jain_2024}.

Among a plethora of LLMs, we opt for the LLaMa 3 8B, a state-of-the-art, ``small'', open-source LLM to generate access control policies from NL requirement specifications \cite{llama3}. 
LLaMa 3 has been pre-trained with more than 15T tokens \cite{llama3}, demonstrating superior performance in popular benchmarks like GSM8K (Math word problem-solving benchmark) \cite{cobbe2021training}, and more importantly HumanEval (Code generation benchmark) \cite{chen2021evaluating}.
The results show that it surpasses other state-of-the-art LMs like Mistral 7B and Gemma 7B significantly in all such benchmarks, especially in code generation from NL, which is similar to access control policy generation \cite{llama3}.
This makes LLaMa 3 the most capable open-source LM at the time of writing \cite{llama3}.
Furthermore, LLaMa 3 adopts Grouped Query Attention \cite{ainslie2023gqa}, which allows LMs to focus on a group of tokens instead of every single token, improving the inference efficiency \cite{llama3}.
Therefore, by choosing LLaMa 3 8B, we ensure the smooth and efficient implementation of \framework, even on low-resource local computers, while improving the reliability of access control policy generation. 
On the other hand, since LLaMa 3 8B is an open-source LLM, it makes \framework{} highly accessible to organizations, especially those with budget constraints \cite{firstpost}.
Also, once \framework{} is deployed within the organization, they do not need to send their confidential access control requirements to an entity that controls a proprietary LLM (like GPT-4) for policy generation.
This avoids security concerns like data leakages caused by providing sensitive information to a third party  \cite{wu2023unveiling, mireshghallah2023can, UbiOps_2024, Dagdelen_Dunn_Lee_Walker_Rosen_Ceder_Persson_Jain_2024}.  

In this research, we use the \emph{``instruct''} version of LLaMa 3 8B, which was created by fine-tuning the LLaMa 3 8B model using the technique called Reinforcement Learning from Human Feedback (RLHF) \cite{ouyang2022training} to follow general user instructions \cite{llama3}.
We further fine-tune it using the access control policy dataset containing NLACPs annotated according to their ACRs (Section \ref{subsec:dataset/access_dataset}) through Parameter Efficient Fine Tuning (PEFT) with low-rank adapters (LoRA) \cite{hu2021lora}. 
While this helps the LLM to be specifically adapted to the access control policy generation domain, it also helps overcome catastrophic forgetting of LM fine-tuning as it only updates a small number of extra parameters of the LM \cite{hu2021lora}. 
In the fine-tuning process, we minimize the negative log-likelihood of the generated access control policy \cite{shen2021generate}.
Once the LLM is fine-tuned, \framework{} uses it to generate access control policies from NLACPs. 

\subsubsection{Step 4.1: Post-processing}
\label{subsubsec:approach/post_processing}

After generating the access control policy as a structured representation of the NLACP, \framework{} then post-processes the policy using the stored information in Step 3. 
This step acts as a validation step to ensure that the generated policy aligns with information pre-defined on the system.
To this end, \framework{} retrieves the most similar entity for each policy component of the access control policy from the vector database and replaces the policy component with it.

\subsection{Step 5: Access Control Policy Verification}
\label{subsec:approach/verification}

Even if an LM (or any ML model) is fine-tuned using domain-specific datasets to generate access control policies from NLACPs, it might sometimes fail to generate the correct policies \cite{skreta2023errors, ni2023lever, heaps2021access}.
This could stem from various factors, such as hallucinations \cite{ni2023lever, shen2021generate}, the complexities and ambiguities of the NLACPs \cite{narouei2018automatic, heaps2021access}, or due to the significant difference between the NL access control requirement and the access control policy \cite{shen2021generate}.
As a solution, once the access control policy is generated, it should be verified to (1) check whether or not it is correct \cite{skreta2023errors, cobbe2021training, ni2023lever} (2) refine the policy automatically if it is incorrect, and (3) alert administrators about errors of the generated policy (if it cannot be refined automatically), allowing them to refine it before adding it to the system \cite{xu2017system}. 
Despite being crucial for a reliable access control policy generation, none of the existing frameworks conduct this verification and refinement step, leaving a significant gap in access control policy generation research \cite{xia2022automated, heaps2021access, slankas2013access, slankas2014relation, narouei2015automatic, narouei2018automatic}.
Therefore, by bridging this gap, \framework{} first introduces a verifier to identify the \emph{incorrectly} generated access control policy based on the joint representation of the NLACP and its structured representation \cite{ni2023lever} and provide feedback.

We develop our verifier as a multi-class sequence pair classifier that identifies correct and incorrect policies with reasons, improving the explainability of our system. 
It outputs a probability distribution over the $m$ classes when the joint representation of the NLACP, and the access control policy generated in Step 4 is provided as the input.
To this end, instead of directly providing the generated policy as part of the input, \framework{} reconstructs the policy as an NL sentence using a pre-defined template. 
For example, consider a simple NLACP \emph{``The doctor is allowed to read records.''}.
Once its access control policy is generated as \emph{[\{`decision': `allow', `subject': `doctor', `resource': `record', `purpose': `none', `condition': 'none'\}]}, it will be transformed into \emph{\textbf{doctor} can \textbf{read} \textbf{records}} according to the template ``\textbf{\{subject\}} can \textbf{\{action\}} \textbf{\{resource\}}''.
It helps the verifier to identify the semantic differences (i.e., errors) between the NLACP and the generated policy easily, as the generated policy is now expressed as an NL sentence (similar to the NLACP) rather than in a structured \emph{``code-like''} format \cite{shen2021generate, skreta2023errors}. 

Then, the reconstructed policy and its NLACP are provided to the verifier to decide whether the policy represents its NLACP (i.e., correct access control policy) or, if not, what makes the policy incorrect according to 11 error types (i.e., incorrect decision, incorrect subject, missing subject, incorrect resource, missing resource, incorrect purpose, missing purpose, incorrect condition, missing condition, incorrect action, or missing ACRs).

Inspired by the previous research \cite{ni2023lever, shen2021generate}, we use BART (Bidirectional and Auto-Regressive Transformers) LM \cite{lewis2019bart, shen2021generate} as the base model for our verifier. 
BART has been successfully used in sequence pair classification tasks such as textual entailment \cite{lewis2019bart, williams2017broad}.
As a result, it has also demonstrated high performance in verification tasks such as Python code verification \cite{li2022making} and mathematical expressions verification \cite{shen2021generate}, which are similar to access control policy verification.

To utilize BART for access control policy verification, first, we add an FFN with two linear layers on top of the final layer hidden states of the last decoder token, \emph{[eos]} of BART \cite{shen2021generate}, to output the classification results. 
Then, we fine-tune it using the policy verification dataset (Section \ref{subsec:dataset/ver_dataset}) by minimizing the cross-entropy loss between the classifier output and the ground truth label \cite{shen2021generate}.
After training the verifier, at the inference time, it outputs the class that represents the correct access control policy if the NLACP and the generated policy match. 
Otherwise, it outputs a reason for the policy being incorrect (i.e., the error type) as feedback, which will then be used in Step 6 to refine the incorrect policy before adding it to the authorization system.


\subsection{Step 6: Iterative Refinement}
\label{subsec:approach/refinement}

After the verifier identifies the error category (if the generated policy is incorrect), it will be converted into an instruction (i.e., \emph{``refinement instruction''} shown in Appendix \ref{appendix:refinement}) that asks the policy generation module to correct the error and re-output the refined policy.
\framework{} repeats this automatic verification-generation process at most $n$ times until the verification result changes into ``correct'' (i.e., the correct policy for the given NLACP is generated). 
If the verification result still indicates the refined policy is incorrect even after $n$ iterations, the final refined policy and the error category are sent to the administrator as feedback. 
It will help the administrator to identify the error of the incorrect policy and refine it manually before adding it to the authorization system \cite{xu2017system}.

\section{Datasets}
\label{sec:datasets}

As we described, we fine-tune LMs utilized in \framework{} using domain-related datasets to improve the reliability of the process \cite{narouei2018automatic}.
This section presents those datasets, namely the \textbf{access control policy dataset} (Section \ref{subsec:dataset/access_dataset}) used to train and evaluate the NLACP identification and policy generation modules, \textbf{access control policy verification dataset} (Section \ref{subsec:dataset/ver_dataset}) used to train and evaluate the verifier, and \textbf{access control policy refinement dataset} (Section \ref{subsec:dataset/refinement_dataset}) used to train the policy generation module to refine incorrectly generated policies. 



\subsection{Access Control Policy Dataset}
\label{subsec:dataset/access_dataset}

While the advancements in ML/NLP domain \cite{vaswani2017attention} help develop reliable automated access control policy generation frameworks, effective adaptation of those advancements is restricted by the lack of domain-related datasets \cite{xia2022automated, alohaly2019automated}.
The main reason for this gap is the confidential nature of organizations' high-level access requirements, except for a few projects that have made their access requirements public \cite{alohaly2019automated}.
Therefore, to address this gap, we developed a synthetic access control policy dataset based on real-world high-level requirement specification datasets for fine-tuning LLM \cite{slankas2014relation}.

\begin{table}[]
    \centering
    \caption{Access control policy dataset containing data from domains like $^{\diamond}$ conference management, $^{\star}$ healthcare, $^{\dagger}$ education, and $^{\bullet}$ the combined dataset from Xiao et al. \cite{xia2022automated}.} 
    \label{tab:acp_dataset}
    \begin{tabular}{ 
   >{\raggedright\arraybackslash}m{3.2cm} 
   >{\raggedleft\arraybackslash}m{1.9cm}
   >{\raggedleft\arraybackslash}m{1.0cm}
   >{\raggedleft\arraybackslash}m{0.7cm}
   }
        \toprule
        {\textbf{Dataset}} &  {\textbf{non-NLACP}} &  {\textbf{NLACP}} &  
        {\textbf{ACRs}}\\
        \midrule
            \multicolumn{4}{c}{\emph{\textbf{Cleaned document-folds (1523 sentences) \cite{slankas2014relation}}}} \\
          Collected ACP (CACP)$^{\bullet}$  & 28    & 112 & 209         \\
          CyberChair (CC)$^{\diamond}$       & 106	& 104 & 157        \\
          IBM (IBM)$^{\dagger}$              & 86    & 114 & 158          \\
          iTrust T2P (T2P)$^{\star}$       & 48    & 341 & 594         \\
          iTrust ACRE (ACRE)$^{\star}$     & 151   & 433 & 863         \\
          \hdashline
          Synthetic data & 522 & 1011 & 2176 \\
          
         \hdashline
         \multicolumn{4}{c}{\emph{\textbf{Overall (3056 sentences)}}} \\
         - Train                & 749   & 1848 & 3677      \\
         - Test                 & 192   & 267  & 481        \\

         \bottomrule
         
    \end{tabular}
    
\end{table}

The real-world datasets are first introduced by Xiao et al. \cite{xiao2012automated} and Slankas et al. \cite{slankas2014relation}, containing sentences from high-level requirement specifications from three systems, namely iTrust, IBM course registration, and the CyberChair conference management. 
Each sentence is first labeled NLACP (1) or non-NLACP (0).
We call these annotations ``sentence type annotations''.
Then, each NLACP sentence is divided into its ACRs and annotated by representing only three policy components: subject, action, and resource.
We call those annotations ``entity annotations''.
However, not only do the provided entity annotations not contain the access decision of each ACR of the NLACP \cite{xia2022automated}, but they also do not represent policy components that provide contextual information to the NLACP, such as purposes and conditions \cite{brodie2006empirical}.
As a result, the provided entity annotations contradict the definitions provided by previous research \cite{brodie2006empirical, yang2021purext}.
For example, according to Slankas et al.'s annotations, the NLACP \emph{``The organisation may use email addresses to answer inquiries.''} 
has two sets of entity annotations representing two ACRs: \emph{``[Organisation]$_{subject}$ may [use]$_{action}$ [email addresses]$_{resource}$''} and \emph{``[Organisation]$_{subject}$ [answer]$_{action}$ [inquiries]$_{resource}$''}.
However, the second set of annotations is incorrect as \emph{``answer inquiries''} should be a purpose, not a separate ACR \cite{brodie2006empirical, yang2021purext}.
Instead, the correct annotation should contain one set of annotations representing only one ACR with subject \emph{``organisation''}, action \emph{``use''}, resource \emph{``email address''}, and purpose \emph{``answer inquiries''}.

Therefore, to begin with, 
we cleaned the datasets (e.g., deduplication) and updated their annotations with the access decisions, purposes, and conditions according to the previous research \cite{yang2021purext, brodie2006empirical}.
Then, we utilized two popular synthetic data generation techniques to generate more diverse access control policies in two steps to better adapt LMs for access control policy generation while minimizing the data scarcity of the domain \cite{narouei2018automatic}. 
They are (1) data augmentation \cite{xia2022automated} and (2) LLM-based synthetic data generation \cite{schmidhuber2024llm, long2024llms}.
More information about the synthetic data generation and annotation can be found in Appendix \ref{appendix:gen_prompt}.

The statistics of the resultant dataset in terms of the number of non-NLACP and NLACP sentences and the number of ACRs are shown in Table \ref{tab:acp_dataset}. 
As shown in Table \ref{tab:acp_dataset}, \emph{Cleaned document-folds} indicate the information about data from five real-world policy documents introduced by Slankas et al. \cite{slankas2014relation} after cleaning and updating annotations. 
Following the previous research \cite{xia2022automated, narouei2015automatic, narouei2018automatic, slankas2014relation}, each fold is specifically used to evaluate our framework, \framework, after training it with the rest of the data \cite{alohaly2019automated}. 
For example, to evaluate \framework{} using data from the dataset CC, the framework is trained using the rest of the folds (i.e., CACP, IBM, T2P, and ACRE), and the synthetic data \cite{alohaly2019automated}.
As a result, even if the fine-tuning involves synthetic data, \framework{} will only be evaluated using real-world access control requirements.
On the other hand, the \emph{``Overall''} dataset is used to train \framework{} and evaluate its overall performance with a diverse set of NLACPs. 
It was created by combining all the sentences from document folds as well as the synthetic data and randomly splitting it into train (80\%), validation (10\%), and test (10\%) sets. 

\subsection{Access Control Policy Verification Dataset}
\label{subsec:dataset/ver_dataset}

The access control policy verification dataset is used to train and evaluate the access control policy verifier discussed in Section \ref{subsec:approach/verification}.
Therefore, the verification dataset should contain sequence pairs representing NLACP and its correct access control policy (i.e., negative sequence pairs) as well as pairs containing NLACP and its incorrect access control policy (i.e., positive sequence pairs).
Since we have the entity annotations for each NLACP from the access control policy dataset (Table \ref{tab:acp_dataset}), we already have the correct access control policies for the NLACPs to create the negative sequence pairs.
Therefore, to generate incorrect policies for each NLACP, we augment the correct policies by manipulating their properties according to each error type mentioned in Section \ref{subsec:approach/verification}.

We applied four manipulation techniques to the correct policies to obtain incorrect policies. 
They are \textbf{(a) Decision change}. Randomly change the decision of an ACR from \emph{allow} to \emph{deny} or vice versa. 
\textbf{(b) Policy component replacement}. Randomly replace the value of a policy component (i.e., subject, action, resource, purpose, and condition) with a different value. 
\textbf{(c) Policy component deletion}. Randomly replace a value of a policy component with \emph{``none''}. 
\textbf{(d) ACR deletion}. If the correctly generated/ground-truth access control policy contains more than one ACR, randomly remove an ACR.
As a result, we have incorrect policies for each NLACP to create the positive sequence pairs.

After generating the sequence pairs, we assigned labels 0 to 10 depending on the error type for the positive sequence pairs and label 11 for the negative sequence pairs. 



\subsection{Access Control Policy Refinement Dataset}
\label{subsec:dataset/refinement_dataset}

The access control refinement dataset is used to teach the access control policy generation module how to refine an incorrect policy identified by the access control policy verification step. 
Therefore, each training example of this dataset should contain the NLACP, its incorrectly generated access control policy, the error type of the incorrectly generated policy, and the correct access control policy of the NLACP as the label. 
All the mentioned components of a training example can be derived from the previously described access control policy dataset (Section \ref{subsec:dataset/access_dataset}) and the access control policy verification dataset (Section \ref{subsec:dataset/ver_dataset}). 
For instance, the NLACP and its correct access control policy can be retrieved from the access control policy dataset. 
Also, the incorrect access control policy for the same NLACP and the error type can be retrieved from the access control policy verification dataset. 
The structure of a training example that combines all the mentioned components is shown in Appendix \ref{appendix:refinement}.

Finally, the dataset is combined with the access control policy dataset (Section \ref{subsec:dataset/access_dataset}) and used to fine-tune the access control policy generation module.

\section{Evaluation and Results}
\label{sec:eval_results}

\subsection{Evaluation}
\label{subsec:eval}

After training the components of \framework, as described in Section \ref{sec:approach}, we evaluate their reliability using the F1 score (i.e., the harmonic mean of the precision and recall \cite{lipton2014optimal}). 
Particularly, we evaluate the \framework's NLACP identification (Section \ref{subsubsec:results/nlacp_id}) and access control policy generation (with iterative refinement) (Section \ref{subsubsec:results/acp_gen}) performance on each document-fold and the test set of the \emph{``Overall''} dataset (Table \ref{tab:acp_dataset}).
Moreover, we evaluate the \framework's reliability in access control policy verification (Section \ref{subsubsec:results/policy_ver}) using the access control policy verification dataset (Section \ref{subsec:dataset/ver_dataset}).



\subsection{Experiment Results}
\label{subsec:results}

This section presents evaluation results of \framework{} and compares them with the existing frameworks \cite{slankas2014relation, slankas2013accessid, narouei2017towards, xia2022automated, narouei2018automatic, narouei2015automatic}.

\begin{table*}[ht]
    \centering
    \caption{The NLACP identification performance of \framework{} in terms of F1 score compared with the existing access control policy generation frameworks. N/R indicates that the results were not reported in the respective paper.}
    \label{tab:acp_id_results}
    \begin{tabular}{ 
   >{\raggedright\arraybackslash}m{3.8cm} 
   >{\centering\arraybackslash}m{1.5cm}
   >{\centering\arraybackslash}m{1.5cm}
   >{\centering\arraybackslash}m{1.5cm}
   >{\centering\arraybackslash}m{1.5cm}
   >{\centering\arraybackslash}m{1.5cm}
   >{\centering\arraybackslash}m{1.5cm}
   >{\centering\arraybackslash}m{1.5cm}}
        \toprule
        {\textbf{Framework}} &  {\textbf{T2P}} &  {\textbf{ACRE}} &  
        {\textbf{IBM}} &  {\textbf{CACP}} & {\textbf{CC}} & {\textbf{Average}} & {\textbf{Overall}}\\
        \midrule
          ACRE \cite{slankas2014relation, slankas2013access, slankas2013accessid} & N/R & N/R & N/R & N/R & N/R & 72\% & - \\
          Narouei et al. \cite{narouei2017towards} & 74.8\% & 83.4\% & 84.4\% & 72.0\% & 73.1\% & 77.5\% & -\\
          Xia et al. \cite{xia2022automated} & \textbf{95.7\%} & 82.2\% & 77.5\% & 89.8\% & 78.5\% & 84.7\% & - \\
          
          \textbf{\framework{}} & 95.2\% & \textbf{91.1\%} & 80.0\% & \textbf{93.2}\% & \textbf{79.6\%} & \textbf{87.9\%} &\textbf{91.9\%}\\


         \bottomrule
         
    \end{tabular}
    
\end{table*}

\begin{table*}[ht]
    \centering
    \caption{The access control policy component extraction performance of \framework{} in terms of the F1 score compared with the existing frameworks. $^{\diamond}$ Extracts subjects, actions, and resources. $^{\star}$ Extracts access decisions, subjects, actions, resources, purposes, and conditions. $^{\dagger}$ With iterative refinement.
    }
    \label{tab:acp_ex_results}
    \begin{tabular}{ 
   >{\raggedright\arraybackslash}m{3.8cm} 
   >{\centering\arraybackslash}m{1.5cm}
   >{\centering\arraybackslash}m{1.5cm}
   >{\centering\arraybackslash}m{1.5cm}
   >{\centering\arraybackslash}m{1.5cm}
   >{\centering\arraybackslash}m{1.5cm}
   >{\centering\arraybackslash}m{1.5cm}
   >{\centering\arraybackslash}m{1.5cm}}
        \toprule
        {\textbf{Framework}} &  {\textbf{T2P}} &  {\textbf{ACRE}} &  
        {\textbf{IBM}} &  {\textbf{CACP}} & {\textbf{CC}} & {\textbf{Average}} & {\textbf{Overall}}\\
        \midrule
          Narouei et al. \cite{narouei2017towards, narouei2018automatic, narouei2015automatic}$^{\diamond }$ & 12.4\% & 16.3\% & 16.8\% &	26.6\% & 12.6\% & 16.9\% & -\\
          Xia et al. \cite{xia2022automated}$^{\diamond}$ & 48.2\% & 34.8\% & 44.7\% & 50.9\% & 29.2\% & 41.6\% & -\\
          
          \textbf{\framework{} (SAR)$^{\diamond}$} & \textbf{87.8\%} & \textbf{72.4\%} & \textbf{87.1\%} & \textbf{87.4\%} & \textbf{73.0\%} & \textbf{81.5\%} & -\\
          \textbf{\framework{} (DSARCP)$^{\star \dagger}$} & \textbf{87.0\%} & \textbf{73.0\%} & \textbf{86.2\%} & \textbf{86.9}\% & \textbf{70.4\%} & \textbf{80.7\%} & \textbf{85.2\%}\\


         \bottomrule
         
    \end{tabular}
    
\end{table*}



         
    

\begin{table*}[ht]
    \centering
    \caption{The access control rule generation performance of \framework{} in terms of F1 score, based on its ability to generate correct ACRs from NLACPs.}
    \label{tab:acp_rule_results}
    \begin{tabular}{ 
   >{\raggedright\arraybackslash}m{3.8cm} 
   >{\centering\arraybackslash}m{1.5cm}
   >{\centering\arraybackslash}m{1.5cm}
   >{\centering\arraybackslash}m{1.5cm}
   >{\centering\arraybackslash}m{1.5cm}
   >{\centering\arraybackslash}m{1.5cm}
   >{\centering\arraybackslash}m{1.5cm}
   >{\centering\arraybackslash}m{1.5cm}}
        \toprule
        {\textbf{Framework}} &  {\textbf{T2P}} &  {\textbf{ACRE}} &  {\textbf{IBM}} &  {\textbf{CACP}} & {\textbf{CC}} & {\textbf{Average}} & {\textbf{Overall}}\\
        \midrule
          \textbf{\framework} & 85.3\% & 69.4\% & 84.9\% & 81.2\% & 70.5\% & 78.3\% & 80.6\%\\


         \bottomrule
         
    \end{tabular}
    
\end{table*}

\begin{table*}[ht]
    \centering
    \caption{The performance of the verifier in identifying correct and incorrect policies. ACR, D, S, A, R, C, and P denote access control rule, decision, subject, action, resource, condition, and purpose, respectively.}
    \label{tab:acp_verification_results}
    \begin{tabular}{ 
   >{\raggedright\arraybackslash}m{1.4cm} 
   >{\centering\arraybackslash}m{1.5cm}
   >{\centering\arraybackslash}m{0.8cm}
   >{\centering\arraybackslash}m{0.8cm}
   >{\centering\arraybackslash}m{0.8cm}
   >{\centering\arraybackslash}m{0.8cm}
   >{\centering\arraybackslash}m{0.8cm}
   >{\centering\arraybackslash}m{0.8cm}
   >{\raggedleft\arraybackslash}m{1.3cm}
   >{\centering\arraybackslash}m{0.8cm}
   >{\centering\arraybackslash}m{0.8cm}
   >{\centering\arraybackslash}m{0.8cm}
   >{\centering\arraybackslash}m{0.8cm}}
        \toprule
         {\textbf{Metric}} &  {\textbf{Correct}} &  \multicolumn{6}{c}{\textbf{Incorrect}} & \multicolumn{5}{c}{\textbf{Missing}} \\
          \cmidrule{3-13}
        & & {\textbf{D}} &  {\textbf{S}} & {\textbf{A}} & {\textbf{R}} & {\textbf{P}} & {\textbf{C}} & {\textbf{ACR}} & {\textbf{S}}  & {\textbf{R}}  & {\textbf{P}}  & {\textbf{C}}\\
        \midrule
          F1 score & & $98\%$ & $98\%$ & $99\%$ & 99\% & 100\% & 99\% & 90\% & 100\% & 100\% & 95\% & 93\%\\
          \cmidrule{3-13}
          & 83\% & \multicolumn{11}{c}{99\%}\\
          \midrule
          Accuracy & \multicolumn{12}{c}{$97\%$}\\
          \hdashline
          F1$_{Macro}$ & \multicolumn{12}{c}{$91\%$}\\
          \hdashline
          F1$_{Weighted}$ & \multicolumn{12}{c}{$97\%$}\\


         \bottomrule
         
    \end{tabular}
    
\end{table*}

\subsubsection{NLACP Identification}
\label{subsubsec:results/nlacp_id}

We evaluate the performance of \framework{} in identifying NLACPs using both document folds and the \emph{``Overall''} datasets introduced in Section \ref{subsec:dataset/access_dataset}. 
The obtained evaluation results compared with the existing access control policy generation frameworks in terms of F1-score are shown in Table \ref{tab:acp_id_results}. According to Table \ref{tab:acp_id_results}, \framework{} achieves an average document fold F1 score of 87.9\%, outperforming the existing frameworks in almost all the document folds.
Moreover, it attains the F1 score of 91.9\% on the test set of the \emph{``Overall''} dataset.

\subsubsection{Access Control Policy Generation}
\label{subsubsec:results/acp_gen}

We evaluate the access control policy generation performance of \framework{} in terms of \emph{access control policy component extraction} and \emph{access control rule generation}.
\emph{Access control policy component extraction} focuses on extracting individual policy components for each ``action'' (i.e., subject, resource, purpose, and condition) of an NLACP \cite{xia2022automated, narouei2015automatic, narouei2018automatic, heaps2021access, slankas2013access, slankas2014relation}.
On the other hand, \emph{access control rule generation} focuses on translating the ACRs of an NLACP into a structured representation, as mentioned in Section \ref{subsec:approach/generation}.

The obtained results for access control policy component extraction in terms of F1 score are reported in Table \ref{tab:acp_ex_results} under two settings of \framework{}: SAR and DSARCP. 
In the SAR setting, we only consider subjects (S) and resources (R) for a given action (A) to calculate the F1 score, enabling the comparison with prior research \cite{xia2022automated, slankas2014relation, narouei2015automatic, narouei2018automatic, slankas2013access}.
In the DSARCP setting, we include access decisions (D), purposes (P), and conditions (C) in addition to subjects and resources when computing the F1 score for each action of NLACPs.

However, we cannot directly compare the performance of \framework{} in extracting policy components with results reported in previous research.
That is because, when developing our document folds, we updated the entity annotations by correcting the errors made by Slankas et al. (Section \ref{sec:datasets}), even if the NLACPs and sentence annotations remained the same.
Therefore, we first implement the ACR extraction modules of two most reliable policy generation frameworks \cite{xia2022automated, narouei2018automatic} that provide sufficient details to implement them from scratch without a significant effort\footnote{We attempted to contact the authors regarding the implementation of their frameworks. Unfortunately, we did not receive any replies.}.
Notably, since both the considered ACR extraction modules utilize general-purpose SRL models without any domain adaptation, we do not need to train them to extract policy components \cite{xia2022automated, narouei2018automatic, narouei2015automatic}.
That being the case, once implemented, we evaluate the performance of those modules on our updated dataset and compare them with \framework{} as shown in Table \ref{tab:acp_ex_results}.

As shown in Table \ref{tab:acp_ex_results}, \framework{} achieves an average document-fold F1 score of 81.5\% in the SAR setting, outperforming all the existing policy generation frameworks significantly by at least 39.9\%. 
Even when considering access decisions and complex components such as purposes and conditions \cite{brodie2006empirical} of each ACR, \framework{} still surpasses the existing frameworks by at least 39.1\% in the DSARCP setting, achieving an average F1 score of 80.7\%.
Furthermore, \framework{} achieves the F1 score of 85.2\% on the \emph{``Overall''} dataset.


Nevertheless, the policy generation framework's ability to extract individual policy components does not accurately evaluate its performance in generating reliable access control rules from NLACPs.
For example, consider an NLACP \emph{``The doctor can read records.''}.
Suppose the access control policy generation framework was able to correctly identify the subject as \emph{``doctor''} but fails to identify the resource as \emph{``records''} for the action \emph{``read''}.
Then, according to previous research \cite{xia2022automated, narouei2015automatic, narouei2018automatic}, there is one true positive (i.e., the subject), no false positives, and one false negative (i.e., for predicting that the \emph{records} is not a policy component).
As a result, the F1 score would be 0.67, indicating that the generated ACR is \emph{partially} correct.
However, in reality, the F1 score should be 0, as the generated ACR is incorrect due to the absence of the correct resource, which may result in access control failures.
Therefore, we further evaluate the performance of \framework{} in terms of its ability to extract/generate ACRs correctly rather than extracting individual components.
The result from Table \ref{tab:acp_rule_results} achieved through the mentioned evaluation technique shows that \framework{} is able to generate ACRs with an average F1 score of 78.3\% across all the document folds and with an F1 score of 80.6\% over the \emph{``Overall''} dataset. 

\subsubsection{Access Control Policy Verification}
\label{subsubsec:results/policy_ver}

To evaluate the verification performance of the \framework's verifier, we utilized the \emph{``Overall''} dataset shown in Table \ref{tab:acp_dataset}, as it contains different types of NLACPs from different domains (e.g., healthcare, education, etc.).
After creating the verification dataset based on the \emph{``Overall''} dataset according to Section \ref{subsec:dataset/ver_dataset}, we first divided it into train (80\%), test (10\%) and validation sets (10\%) randomly \cite{meng2020exploring}. 
Then, we train the verifier using the training set, select the best model/checkpoint based on the validation set, and finally, evaluate the verifier's performance in the test set. 
The obtained evaluation results are shown in Table \ref{tab:acp_verification_results} in terms of average accuracy and F1 scores. 

According to Table \ref{tab:acp_verification_results}, \framework's verifier is able to identify incorrect and correct access control policies with F1 scores of 99\% and 83\%, respectively, achieving an overall macro F1 score of 91\%. 
It further shows that the verifier identifies each error type with an F1 score of more than 90\%.

According to Section \ref{subsubsec:results/nlacp_id}, \ref{subsubsec:results/acp_gen}, and \ref{subsubsec:results/policy_ver}, the evaluation of \framework{} on five real-world datasets show that it outperforms the current state-of-the-art, significantly improving the access control policy generation reliability.
However, one might argue that these results may not clearly capture \framework's performance in real-world scenarios, as the datasets were already processed by the original authors of the datasets \cite{slankas2014relation}.
Therefore, we also analyzed the performance of \framework, in a real-world policy generation scenario and reported our findings in Appendix \ref{sec: usecase}.

\section{Discussion}
\label{sec:discussion}

We propose a novel retrieval-based access control policy generation framework, \framework{} as shown in Figure \ref{fig:framework}, innovatively adapting pre-trained open-source LMs to the access control policy generation domain. 
To the best of our knowledge, this is the first framework that utilizes LMs in identifying NLACPs, translating them into access control policies through RAG, and finally, verifying and refining the generated policies automatically using our novel iterative verification-refinement mechanism.
Our evaluation suggests that \framework{} outperforms all the existing state-of-the-art policy generation frameworks (by 3.2\% in NLACP identification and by 39.1\% in access control policy component extraction),
significantly improving the reliability of access control policy generation \cite{narouei2015towards, xia2022automated, slankas2013access, slankas2014relation, narouei2015automatic, narouei2018automatic}. 
In this section, we discuss those evaluation results to understand why \framework{} performs better than existing frameworks and why it is vital in access control policy generation.

\textbf{NLACP Identification.} 
The superior results attained by \framework{} in NLACP identification (average F1 score of 87.9\%) according to Table \ref{tab:acp_id_results} suggest that it identifies access requirements buried in high-level requirement specification documents better than the existing frameworks.
Therefore, compared to the existing frameworks, the probability of access control failures due to missing policies will be reduced \cite{slankas2013access, slankas2013accessid, slankas2014relation, narouei2015automatic, narouei2018automatic, narouei2017towards, xia2022automated, xiao2012automated}. 
This improvement is mainly due to the capability of \framework's BERT-based NLACP identification module to learn language representations \cite{devlin2018bert}.
As a bi-directional LM, BERT attends tokens from both sides of a given token with its ``self-attention'' mechanism \cite{devlin2018bert}, learning to incorporate context from each word of the sentences when classifying them \cite{patel2022bidirectional}.
Consequently, BERT can identify NLACPs more accurately compared to other traditional ML techniques used in existing frameworks, such as SVM and k-NN \cite{slankas2014relation, slankas2013access, slankas2012classifying, slankas2013accessid, narouei2017towards}, improving the reliability of \framework.

\textbf{Access Control Policy Generation.} 
As we reported in Section \ref{subsubsec:results/acp_gen}, \framework{} outperforms all the existing frameworks in access control policy generation significantly \cite{xia2022automated, narouei2015automatic, narouei2018automatic, narouei2017towards}.
 \framework{} excels in extracting not only basic components like subjects, actions, and resources but also purposes and conditions when generating policies, achieving an average F1 score of 80.7\%.
 It shows that even if components with complex linguistic forms like purposes and conditions  \cite{brodie2006empirical} are involved in policy generation, \framework{} performs with high reliability, which cannot be achieved in any existing framework.
As a result, in contrast to existing frameworks, \framework{} allows organizations to generate access control policies that ensure access to resources is granted or denied based on legitimate reasons and contextual factors, thereby enhancing security and compliance.

Furthermore, including domain/organization-specific information as a non-parametric memory aids \framework{} to generate access control policies that match the pre-defined entities like users and resources, as shown in Figure \ref{fig:ablation/retrieval}.
Utilizing such information provides \framework{} contextual awareness necessary for generating policies accurately in unknown domains \cite{ding2024survey, han2023comprehensive}, which is lacking in existing frameworks.
As a result, \framework{} produces correct access control policies compared to the current state-of-the-art, especially when evaluated using out-of-domain data from document-folds, achieving superior F1 scores as shown in Table \ref{tab:acp_ex_results}.
Also, unlike in existing frameworks, this removes the need to train \framework{} repetitively with the organization's confidential access requirements to adapt it to specific contexts.
The significance of incorporating organization-specific information is further evaluated in our ablation study reported in Appendix \ref{subsec:ablation/retrieval}.

Even if the organization-specific entities are not provided, our ablation study reported in Appendix \ref{subsec:ablation/retrieval} shows that \framework{} still generates access control policies with significantly high reliability compared to the existing frameworks \cite{xia2022automated, narouei2015automatic, narouei2018automatic, slankas2014relation}.
This proves that \framework{} can also be used to develop the initial set of access control policies of an organization reliably, where there are no pre-defined entities \cite{narouei2018automatic}.
However, in that case, the existing state-of-the-art frameworks \cite{xia2022automated, narouei2018automatic}, often extract policy components for actions that are not related to the current access control policy \cite{narouei2015towards, narouei2018automatic, narouei2015automatic, xia2022automated}, leading to low reliability. 
For example, consider the NLACP from CACP dataset, \emph{``The organisation may use email addresses to answer inquiries.''} that has on ACR with the subject \emph{``organisation''}, action \emph{``use''}, resource \emph{``email address''}, and purpose \emph{``answer inquiries''} \cite{yang2021purext, brodie2006empirical}.
However, when that NLACP is given to any existing policy generation framework, it will identify two ACRs as \emph{``The organisation may use email addresses''} and \emph{``The organisation answer inquiries''}.
Consequently, the extracted components from the ACR with the action \emph{``answer''} will be considered as false positives \cite{yang2021purext}.
Even though Xia et al. used a ``domain dictionary'' \cite{xiao2012automated} to store such ``unrelated actions'' \cite{xia2022automated} and filter out unwanted ACRs, they are not flexible enough to handle all such cases.
For instance, if the action \emph{``answer''} is included in that dictionary to filter out the unwanted ACR detected from the aforementioned NLACP, it might reject a valid ACR with the action \emph{``answer''} in another access control domain.
Therefore, those false positives are harder to avoid, resulting in low precision, leading to significantly lower F1 scores \cite{narouei2018automatic} compared to \framework{}\footnote{We implemented the existing frameworks like Xia et al.'s framework \cite{xia2022automated} as mentioned in Section \ref{subsubsec:results/acp_gen}, assuming that all the information about their implementations, such as the entire ``domain dictionary'', is reported in the paper.
However, the authors may have reported only a \emph{portion} of the dictionary, partly affecting the reliability of Xia et al.'s framework.}.

Nonetheless, as we explained in Section \ref{subsec:approach/refinement}, the ML/NLP-based policy generation framework may still generate incorrect policies due to reasons like complexities and ambiguities of high-level requirement specification (e.g., ambiguous policy components, grammatically incorrect NLACPs) \cite{narouei2015automatic, jayasundaravision, jayasundara2023sok, del2018systematic}.
Therefore, as reported in our ablation study (Appendix \ref{subsec:ablation/refinement}), by iteratively identifying and refining such errors automatically, \framework{} improves the reliability of access control policy generation even further compared to existing frameworks.

\textbf{Access Control Policy Verification.} 
In order to refine the generated policies, incorrect policies, as well as the error category of the incorrect policies, should be identified correctly by the access control policy verifier. 
That is because \framework{} should only refine the incorrect policies based on the specific error category \cite{kambhampati2024llms}.
As reported in Section \ref{subsubsec:results/policy_ver}, \framework's BART-based verifier fulfills the mentioned requirements by identifying incorrect policies with an F1 score of 99\% and error categories with an F1 score of more than 90\%.
Therefore, the correct and necessary information required to refine an incorrect policy will be provided to the policy generation module, leading to the correct policy after refinement (assuming the policy generation module refines the policy correctly).

However, as we found out, the verifier sometimes misidentifies correct access control policies as incorrect (i.e., false positives), often saying that the generated policy is missing an ACR.
Consequently, the F1 score for identifying missing ACR becomes relatively lower than F1 scores for identifying other error categories due to low precision (Table \ref{tab:acp_verification_results}).
In such cases, interestingly, even though the refinement instruction asks the policy generation module to address the error \emph{``missing ACR''} (i.e., incorrect verification result), the policy generation module keeps outputting the correct policy generated in the first place without any modification.
That is because the policy generation module does not address an error that is unavailable in the generated policy.
It suggests that, even if the BART-based verifier sometimes misclassified the policies, the policy generation module based on LLaMa 3 (a more advanced LLM compared to BART) rectifies the effect of that misclassification. 


\subsection{Limitations}

In this section, we point out the limitations of \framework{} and highlight future improvements.

\textbf{Policy quality assurance.} To ensure the quality of access control policies, we have to make sure that they are free from conflicts (i.e., \emph{consistency}), they match the high-level requirements/goals (i.e., \emph{correctness}), they are not redundant (i.e., \emph{minimality}), they address actions only relevant to the user (i.e., \emph{relevance}), and they cover all possible access request scenarios (i.e., \emph{completeness}) \cite{bertino2018challenge}.
While \framework{} ensures the \emph{correctness} of the generated policies via verification and two refinement stages (i.e., automatic and manual), it can further be extended to facilitate the other four requirements. 
To this end, we will use Binary Decision Diagrams (BDD) \cite{jabal2019methods, morisset2019framework} to improve \framework{} further, so that it can evaluate the generated policies in the end to ensure their \emph{consistency, minimality, relevance, and completeness} as a future work.


\textbf{Detailed and necessary contextual information.} In this research, we showed that utilizing simple pre-defined entities like subjects and resources as the context would significantly improve the reliability of policy generation. 
However, more detailed information about the organization can be incorporated to make the context more meaningful than individual entities, improving RAG \cite{ding2024survey}.
Therefore, \framework{} can further be improved to process and use detailed information like organizational breakdown structure (OBS) as the context, potentially improving the policy generation reliability. 

Furthermore, \framework{} retrieves $k$ (a constant) most similar entities to the NLACP when generating policies through RAG.
However, all the retrieved entities may not be required to generate a policy from a particular NLACP.
For example, to translate the NLACP \emph{``The doctor can read records.''} \framework{} retrieves $k=5$ subjects even though the NLACP has only one subject.
This irrelevant information may sometimes mislead the LLM to generate incorrect policies \cite{yan2024corrective}.
Therefore, we will further improve \framework{} to measure the relevancy of retrieved information using an access control-specific retrieval evaluator \cite{yan2024corrective} and use only the relevant $j$ (a variable) pieces of information to generate the policies.

\section{Conclusion and Future Works}
\label{sec:conclusion}

In this research, we propose \framework, a retrieval-based access control policy generation framework. 
It is designed to help system administrators reliably generate access control policies from high-level requirement specification documents, reducing human errors.
Utilizing ``small'' open-source transformer-based LMs, \framework{} enables its efficient and local deployment within the organization, ensuring the confidentiality of the access control policy generation process.
In contrast to existing frameworks, \framework{} identifies and translates complex access control requirements containing multiple ACRs with even intricate components like purposes and conditions, in addition to subjects, actions, and resources.
We showed that incorporating organization-specific information like subjects and resources pre-defined in the organization provides essential contextual information in policy generation, improving \framework's reliability significantly.
We further demonstrate \framework's ability to successfully extract hidden ACRs from NLACPs and generate access control policies by incorporating the access decisions of each ACR through a real-world application.
More importantly, \framework{} introduces a novel iterative verification-refinement mechanism to correct its incorrect generations automatically.
While it bridges a significant gap in existing access control policy generation frameworks, it also improves the correctness of the generated policies and, in turn, improves their quality \cite{bertino2018challenge}.
Finally, if the automatic refinement fails, \framework{} provides feedback to administrators, helping them to refine incorrect policies manually.
Additionally, we also release the annotated datasets used to train and evaluate \framework, addressing the data scarcity in the domain \cite{ narouei2018automatic, xia2022automated}.

As future works, we will first improve the feedback that \framework{} provides to the administrator, providing information such as instructions to refine the incorrect policy \cite{xu2017system}.
To this end, we plan to design the feedback according to explainable security (XSec) concepts \cite{vigano2020explainable}, involving system administrators through user studies \cite{schuler1993participatory, holtzblatt1997contextual}.

Secondly, we will develop a usable policy configuration interface according to Nielsen's usability quality components \cite{nielsen1994usability} to present the policy generation framework to administrators. 
Subsequently, we will use contextual design \cite{holtzblatt1997contextual} to involve system administrators to improve its usability further iteratively \cite{schuler1993participatory}. 
Then, we will empirically evaluate the usability and reliability of our access control policy generation tool through lab study, allowing the participants to perform policy generation tasks using the tool \cite{brodie2005usable, brodie2006empirical}. 
Finally, we will evaluate their subjective satisfaction with our tool using standard evaluation instruments such as the System Usability Scale (SUS) \cite{brooke1996sus} or Post-Study System Usability Questionnaire (PSSUQ) \cite{fruhling2005assessing}.


\section*{Ethics Considerations and Compliance with the Open Science Policy}

In this research, we introduce a \framework{}, novel retrieval-based access control policy generation framework based on transformer-based LMs. 
It helps system administrators generate access control policies from high-level requirement specifications with low overhead, reducing access control failures due to human errors.

First, when choosing the LMs/LLMs for our research, we focused on ``small'' open-source LMs rather than utilizing powerful proprietary LLMs like GPT-4.
That is because closed-source GPT-4-like LMs are exclusively controlled by a private entity \cite{Dagdelen_Dunn_Lee_Walker_Rosen_Ceder_Persson_Jain_2024}.
In that case, the policy generation frameworks based on such LLMs would need to send the confidential NLACPs to a different organization that controls the LLM for generating access control policies, posing security concerns like data leakages \cite{wu2023unveiling, UbiOps_2024, Dagdelen_Dunn_Lee_Walker_Rosen_Ceder_Persson_Jain_2024}.
Furthermore, the entity controlling the LM may change it at any time under their discretion, affecting the reliability of policy generation or revoke access to the LM altogether \cite{Dagdelen_Dunn_Lee_Walker_Rosen_Ceder_Persson_Jain_2024}.
In contrast, by utilizing ``small'', open-source LMs, \framework{} avoids those negative outcomes while improving the reliability of access control policy generation as we show in Section \ref{sec:eval_results}.
It allows the efficient deployment of \framework{} within the organization even in low resource environments, providing the control of \framework{} to the organization.
Therefore, \framework{} does not send an organization's confidential access control requirements to a separate entity, preserving its confidentiality.

Secondly, we adapt the LLMs by fine-tuning them to identify NLACPs, generate access control policies, and verify them (Section \ref{sec:datasets}).
To this end, we only use real-world yet publicly available datasets introduced by Slankas et al. \cite{slankas2014relation}, ensuring that they do not contain any sensitive information like email addresses, phone numbers, etc.
We also generate synthetic data and use them only during the fine-tuning stage of \framework, as mentioned in Section \ref{subsec:dataset/access_dataset}. 
After the fine-tuning process, we evaluate \framework{} using the public real-world dataset mentioned above to assess its performance in identifying real-world NLACPs and generating access control policies from them.
We further showcase \framework's performance in real-world policy generation scenarios by generating access control policies from high-level requirement specifications of the HotCRP.com conference management website as reported in Appendix \ref{sec: usecase}.

Finally, since the existing ML-based automated policy generation approaches are far from 100\% accurate \cite{del2018systematic, xu2017system, xia2022automated, narouei2015automatic, narouei2018automatic}, the chance of applying incorrect policies generated by those frameworks to the authorization system is high.
Consequently, these erroneous policies may cause access control failures, leading to data breaches \cite{Page_2023}.
Therefore, \framework{} attempts to avoid applying such incorrectly generated policies by LMs, using a novel automated iterative verification-refinement mechanism (Section \ref{subsec:approach/refinement}).
This mechanism involves identifying an incorrectly generated policy and re-generating the correct policy by iteratively addressing the identified error using LMs.
If the automatic iterative refinement fails to correct incorrect policies, we also incorporate system administrators as a fail-safe to refine the incorrect policies manually before adding them to the authorization system.
As a result, as we show in our ablation study (Appendix \ref{appendix:refinement}), our proposed verification-refinement mechanism of \framework{} significantly improves the reliability of the fully automated policy generation process, proving that it is a promising approach for reducing access control failures due to incorrect policies.

\subsection*{Availability}
To comply with open science policy, we make our code and datasets used to implement, train, and evaluate \framework{} public via \gitrepo.

\bibliographystyle{plain}
\bibliography{bibtex}

\appendix
\section{Appendix}
\label{appendix}

This section provides additional information that is pertinent to our analysis and methodology.







\subsection{Access Control Policy Generation}
\label{appendix:gen_prompt}

As we mentioned in Section \ref{subsec:dataset/access_dataset}, we used data augmentation and LLM-based synthetic data generation techniques to improve the original dataset introduced by Slankas et al. \cite{slankas2014relation}.
In the data augmentation step, we utilized BERT embeddings to substitute words of the sentences randomly and back translation \cite{xia2022automated}.
For example, by substituting random words of the sentence \emph{``The doctor can read patient's records.''} using BERT embeddings \cite{devlin2018bert}, we can generate a sentence like \emph{``ER doctors can read patient's medical history.''}.
Similarly, in back translation, we translated the original NLACP to German and translated it back to English using \cite{wmt}.
By doing so, we attempted to introduce variations to the sentences while keeping the sentence structure intact, helping LMs to improve their generalization.

After generating synthetic data through data augmentation, we manually annotated them, representing six components: subject, action, resource, purpose, condition, and access decision, separated into ACRs. 
If any of the above six components were absent in the sentence, the component's value was replaced with ``\emph{none}''.
Furthermore, to support NLACP identification, we labeled synthetic non-NLACP sentences as 0 and NLACP sentences as 1, keeping the sentence type annotations from Slankas et al.'s dataset unchanged.

When it comes to LLM-based synthetic data generation, we utilized OpenAI's GPT-4 \cite{achiam2023gpt} through in-context learning.
To this end, we created the following prompt with instructions to generate a synthetic NL access control requirement for a given domain (i.e., healthcare, education, etc.) and its entity annotation according to the provided access control policy definition. 
We also enriched the prompt with few-shot examples to improve the quality of the generated synthetic data following \cite{long2024llms}.

\begin{lstlisting}[]
You are a system administrator working in the <domain> domain who is able to generate domain-related natural language access control requirements used to control access of users to resources in an organization. Generate such a natural language access control requirement and generate its structured representation according to the provided access control policy definition below. Output the natural language policy and its representation according to the following format without providing any additional text.

<Access control policy definition>
<few shot example 1>
<few shot example 2>
<few shot example 3>

Policy: <natural language access control policy>
ACP: <structured representation>

\end{lstlisting}

After generating the synthetic data, we manually reviewed each sentence and its annotation to accept, reject, or refine them if necessary.
It is important to emphasize that we only use such a commercialized black-box LLM to generate \emph{synthetic} data for fine-tuning our LMs. 
We do not use it for the policy generation (Section \ref{subsec:approach/generation}) that deals with organization's confidential real-world access control requirements as it requires sending them to a private entity that controls GPT-4 exclusively, making the organization prone to data leakages \cite{UbiOps_2024, Dagdelen_Dunn_Lee_Walker_Rosen_Ceder_Persson_Jain_2024} (Section \ref{subsec:approach/generation}).

After preparing the access control policy dataset (Section \ref{subsec:dataset/access_dataset}), we convert each annotated sentence into chat messages (i.e., training examples) with three roles: \emph{system} which provides instructions to perform a particular task (Line 1-5), \emph{user} who provides the input (i.e., user query containing the NLACP and relevant information that can be used to translate the NLACP) to the task (Line 7-10), and \emph{assistant} who provides the answer to the user query according to the system instructions (Line 12-14). 
Finally, we applied the following LLaMa 3 chat template to each training example during the training phase of the \framework's access control policy generation module as advised by \cite{llama3modelcard}. 

\begin{lstlisting}[]
<|begin_of_text|><|start_header_id|>system<|end_header_id|>
<Access control policy definition>

Given a natural language sentence (i.e., NLACP), generate an access control policy according to the above Access Control Policy Definition. If a value for any key in any Python dictionary cannot be found in the NLACP, set the value to 'none'. To identify subject, resource, purpose, and condition, use the entities provided as a dictionary, 'Available entities' as an aid. If none of the provided Available entities match the entities of the NLACP or there are no 'Available entities' provided, use your judgment to select the most suitable entity within the NLACP
<|eot_id|>

<|start_header_id|>user<|end_header_id|>
NLACP: <NLACP>
Available entities: <relavant entities>
<|eot_id|>

<|start_header_id|>assistant<|end_header_id|>
<entitiy annotation>
<|eot_id|>
\end{lstlisting}

\textbf{<|start\_header\_id|>x<|end\_header\_id|>} is a special token sequence for LLaMa 3 LLM to indicate the starting points of each role \textbf{x} involved in the chat. 
\textbf{x} can be replaced with either \emph{``system''}, \emph{``user''} or \emph{``assistant''}. 
Additionally, \textbf{<|begin\_of\_text|>} and \textbf{<|eot\_id|>} indicate the beginning of the training example and the end of the turn of each role (i.e., system, user, or assistant), respectively.

\subsection{Access Control Policy Refinement}
\label{appendix:refinement}

As described in Section \ref{subsec:dataset/refinement_dataset}, the complete structure of a training example in the access control policy refinement dataset can be shown as follows.
Similar to training examples from the access control policy dataset, we use the same chat message template to create training examples.
Finally, we applied the LLaMa 3 chat template to each training example (i.e., chat messages) during the training phase of the \framework's access control policy generation module.

\begin{lstlisting}
<|begin_of_text|><|start_header_id|>user<|end_header_id|>
<Access control policy definition>

You generated <incorrectly generated policy> for the sentence, <NLACP> based on the mentioned Access Control Policy Definition. However, the following error is found.

1. <error>

Please address the error and output the corrected access control policy according to the mentioned definition. Think step-by-step to first provide the reasoning steps/thought process. Then after the 

### Corrected: 

provide the corrected policy without any other text.
<|eot_id|>

<|start_header_id|>assistant<|end_header_id|>
<reasoning according to the chain-of-thought>

### Corrected: 
<correct access control policy>
<|eot_id|>
\end{lstlisting}

It is important to note that, instead of providing the ``correct access control policy'' alone as the label (i.e., the response of the ``assistant'') for the training example, we added the intermediate reasoning steps according to \emph{chain-of-thought} \cite{wei2022chain}, before providing the answer/label. 
It helps train the policy generation module to address the provided \texttt{<error>} step-by-step, outputting the intermediate steps that lead to the correct access control policy at the end. 
By doing so, the policy generation module based on LLaMa 3, which is pre-trained to predict the next token given the previous tokens (i.e., causal language modeling) \cite{llama3}, will be able to accurately generate the refined access control policy based on generated the refining steps, in the end, \cite{wei2022chain, kambhampati2024llms}. 

\subsection{Real-World Application}
\label{sec: usecase}

In Section \ref{subsec:results}, we demonstrated the higher reliability of our framework, \framework, compared to the existing frameworks \cite{xia2022automated, narouei2018automatic, slankas2014relation, xiao2012automated}, using the datasets described in Section \ref{sec:datasets}.
The sentences of those datasets were extracted from high-level requirement specification documents using a pre-defined set of grammar by Slankas et al. \cite{slankas2013access}. 
However, Slankas et al. further refined those sentences to align with the techniques used by their policy generation framework \cite{slankas2013access}. 
For example, they replaced shorthand and removed text in the sentences of the datasets that their dependency parser would not recognize when extracting ACRs \cite{slankas2013access}.
Such processing techniques have made the datasets somewhat \emph{``artificial''}.
Consequently, one might argue that the results reported in Section \ref{subsec:results} based on those datasets may not accurately reflect the performance of \framework{} in a real-world setting.
Therefore, to address this concern, we applied \framework{} to the content from the HotCRP conference management website's privacy policy webpage \cite{hotcrp} containing 39 English sentences. 
We opted for \emph{privacy policies} instead of \emph{access control requirements} due to (1) the unavailability of confidential access control requirements of any organization and (2) the similarities between privacy and access control requirements, such as policy components \cite{yang2021purext, brodie2006empirical}.

To begin with, we copied the entire page that describes privacy policies and reformatted it according to the markdown format, for ease of processing and visualization.  
Provided that reformatted document, \framework{} first conducted pre-processing through techniques such as separating paragraphs, coreference resolutions, sentence segmentation, and tokenization, as we described in Section \ref{subsec:approach/preprocessing}. 
After the pre-processing, \framework{} correctly identified NLACPs in the HotCRP privacy policy webpage with an F1 score of 95\%, showcasing its ability of NLACP identification even in a new domain. 
Once a sentence is identified as an NLACP, \framework{} translated it into its structured representation as described in Section \ref{subsubsec:results/acp_gen}.
When generating policies, first, we did not provide the entities like subjects and resources relevant for policy generation through information retrieval.
Consequently, we found that \framework{} sometimes does not identify ambiguous subjects like \emph{``hotcrp.com''} and identify partial resources like \emph{``submission''} instead of \emph{``submission artifact''}.
However, once we provided entities as relevant information to generate policies, \framework{} successfully identified such ambiguous subjects and generated policies that match the pre-defined entities like resources.
This shows the importance of providing the relevant information for policy generation (other than the NLACP), further validating our results obtained in the ablation study (Appendix \ref{subsec:ablation/retrieval}).
Furthermore, we found that \framework{} uncovers complex hidden ACRs within NLACPs and generates access control policies accurately. 
For instance, given the NLACP, \emph{``Demographic data is stored in user global profiles, and can only be modified by users (never by site managers).''}, it generates the structured representation as,
\begin{lstlisting}[language=Python]
    {
        decision: allow; subject: user; action: modify; resource: demographic data; purpose: none; condition: none 
        |
        decision: deny; subject: site manager; action: modify; resource: demographic data; purpose: none; condition: none
    }
\end{lstlisting}
Notably, \framework{} identified that the \emph{``site manager''} cannot \emph{``modify'' ``demographic data''} even though it is not mentioned in the NLACP clearly. 
Consequently, \framework{} achieves the F1 score of 87.3\% in extracting policy components from HotCRP privacy policies\footnote{To get the F1 scores in NLACP identification and policy generation, we conducted the same annotation process mentioned in Appendix \ref{appendix:gen_prompt} to annotate the sentences from HotCRP privacy policy}. 
In contrast, once we provided the same NLACP to the SRL-based Xia et al.'s framework \cite{xia2022automated}, it only identified the subject as \emph{``by users by site managers''} and the resource as \emph{``demographic data''} for the action \emph{``modify''}.
While it did not correctly extract either subject (i.e., \emph{``user''} or \emph{``site manager''}), it cannot identify that the permission for the \emph{``site manager''} to \emph{``modify''} the \emph{``demographic data''} is \emph{\textbf{denied}}.

Finally, we found that the verification-refinement mechanism of \framework{} also contributed to this significantly high policy generation reliability compared to the current state-of-the-art \cite{xia2022automated}.
For example, at first, \framework{} generated the access control policy for the NLACP \emph{``In most cases, The HotCRP.com service does not own the data you provide.''} as,
\begin{lstlisting}
[
    {
        "decision": "allow",
        "subject": "hotcrp.com",
        "action": "own",
        "resource": "data",
        "condition": "none",
        "purpose": "none"
    }
]
\end{lstlisting}
with the decision as \emph{``allow''}, even though it should be \emph{``deny''}.
This error was caught by \framework's verifier, identifying the policy as an incorrect policy with the incorrect decision.
Therefore, based on that verification result, \framework{} refined the policy, leading to the correct access control policy as,
\begin{lstlisting}
[
    {
        "decision": "deny",
        "subject": "hotcrp.com",
        "action": "own",
        "resource": "data",
        "condition": "none",
        "purpose": "none"
    }
]
\end{lstlisting}
within two iterations.

\subsection{Ablation Study}
\label{sec:ablation}

We perform an ablation study for \framework{} and compare the results with the same existing frameworks we used in Section \ref{sec:eval_results}. 

\subsubsection{Effect of Utilizing Organization-specific Information}
\label{subsec:ablation/retrieval}

\begin{figure}[!ht]
    \centering
    \begin{tikzpicture}
        \begin{axis}[
            ybar,
            bar width=.19cm,
            width=1\columnwidth,
            height=0.75\columnwidth,
            legend style={at={(0.45,-0.25)}, anchor=north, legend columns=1, 
            },
            symbolic x coords={T2P, ACRE, IBM, CC, CACP},
            xtick=data,
            nodes near coords,
            every node near coord/.append style={font=\small}, 
            ymin=0,
            xlabel={Dataset},
            ylabel={F1 score (\%)},
            enlarge x limits={abs=0.65cm},
            x=1.5cm,
            grid=both, 
            grid style={line width=.1pt, draw=gray!50}
        ]
        \addplot coordinates {(T2P,12.4) (ACRE,16.3) (IBM,16.8) (CC,12.6) (CACP,26.6)};
        \addplot coordinates {(T2P,48.2) (ACRE,34.8) (IBM,44.7) (CC,29.2) (CACP,50.9)};
        \addplot coordinates {(T2P,80.5) (ACRE,68.3) (IBM,75.7) (CC,52.7) (CACP,71.6)};
        \addplot coordinates {(T2P,84.7) (ACRE,69) (IBM,84.9) (CC,69.4) (CACP,84.4)};
        \addplot coordinates {(T2P,87) (ACRE,73) (IBM,86.2) (CC,70.4) (CACP,86.9)};
        \legend{Narouei et al. \cite{narouei2017towards}, Xia et al. \cite{xia2022automated}, RAgentV (w/o retrieval), RAgentV (w/o post-processing), RAgentV}
        \end{axis}
    \end{tikzpicture}
    \caption{Effect of utilizing organization-specific information to generate access control policies (Section \ref{subsec:approach/generation}) and to post-process generated policies (Section \ref{subsubsec:approach/post_processing}).}
    \label{fig:ablation/retrieval}
\end{figure}
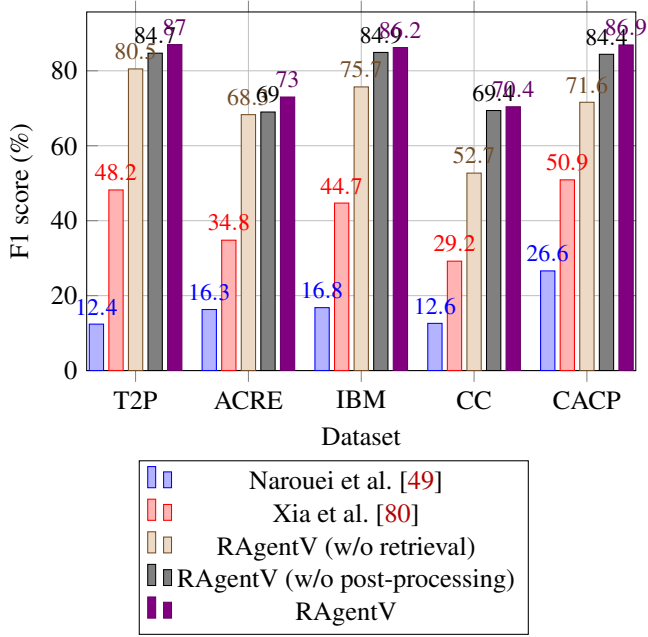

In \framework, we utilize organization-specific entities like subjects and resources not only as an input to generate access control policies from NLACPs but also to post-process the generated policies in Step \ref{subsubsec:approach/post_processing}.
According to Figure \ref{fig:ablation/retrieval}, the reliability of \framework{} drops considerably on all five document-folds when the organization-specific information is not provided to generate access control policies (as an input) through information retrieval (Section \ref{subsec:approach/retrieval}). 
It highlights the importance of organization-specific information when generating the correct enforceable access control policies, despite being the reliability still significantly higher than the current state-of-the-art frameworks (i.e., Narouei et al. \cite{narouei2017towards, narouei2018automatic, narouei2015automatic} and Xia et al. \cite{xia2022automated}) even without that information.

As shown in Figure \ref{fig:ablation/retrieval}, the effect of not providing organization-specific information to generate policies from NLACPs varies across different datasets.
While it causes an absolute performance drop of 17.7\%, 15.3\%, and 10.5\% for CC, CACP, and IBM, respectively, the drop is smaller for T2P (6.5\%) and ACRE (4.7\%).
The main reason for the relatively smaller drop is that both T2P and ACRE datasets were derived from the same iTrust dataset \cite{slankas2014relation}. 
In other words, while T2P and ACRE contain different NLACPs, they often address similar entities like ``LHCP'', ``HCP'', and ``Patient''.
Therefore, as mentioned in Section \ref{subsec:dataset/access_dataset}, in the document-fold evaluation of T2P (ACRE), we train the model with the other folds, including ACRE (T2P).
As a result, the policy generation module can learn some percentage of entities of the T2P (ACRE) dataset through the ACRE (T2P) dataset during its training stage, reducing the necessity of a non-parametric memory (i.e., information about pre-defined entities).
On the other hand, since the origins of IBM, CC, and Collected are different, they seem to benefit significantly from their domain-specific information (10.5\% minimum improvement in F1 score), as indicated by their higher performance drops when that information is not provided to generate policies. 

Furthermore, Figure \ref{fig:ablation/retrieval} shows that the policy generation reliability of \framework{} is slightly reduced if the generated policies are not post-processed using the organization-specific information (Section \ref{subsubsec:approach/post_processing}). 
While that reduction is relatively higher for ACRE (4\%), it is lower for CC (1\%), IBM (1.3\%), T2P (2.3\%), and CACP (2.5\%).
Nonetheless, it shows that even though the organization-specific information is provided as input to the policy generation module to generate policies through RAG, post-processing the generated policies (with the same information) can further help improve the reliability of the process.

\subsubsection{Effect of Iterative Refinement}
\label{subsec:ablation/refinement}

\begin{figure}[!ht]
    \centering
    \begin{tikzpicture}
        \begin{axis}[
            ybar,
            bar width=.19cm,
            width=1\columnwidth,
            height=0.75\columnwidth,
            legend style={at={(0.45,-0.25)}, anchor=north, legend columns=2, 
            /tikz/every even column/.append style={column sep=0.5cm}
            },
            symbolic x coords={T2P, ACRE, IBM, CC, CACP, Overall},
            xtick=data,
            nodes near coords,
            every node near coord/.append style={font=\small}, 
            ymin=0,
            xlabel={Dataset},
            ylabel={F1 score (\%)},
            enlarge x limits={abs=0.65cm},
            x=1.2cm,
            grid=both, 
            grid style={line width=.1pt, draw=gray!50}
        ]
        \addplot coordinates {(T2P,83.7) (ACRE,66) (IBM,84.7) (CC,66.7) (CACP,79.2) (Overall,77.9)};
        \addplot coordinates {(T2P,85.3) (ACRE,69.4) (IBM,84.9) (CC,70.5) (CACP,81.2) (Overall,80.6)};
        \legend{\framework{} (w/o refinement), \framework}
        \end{axis}
    \end{tikzpicture}
    \caption{Effect of iteratively refining the generated policies based on the verification result.}
    \label{fig:ablation/refinement}
\end{figure}
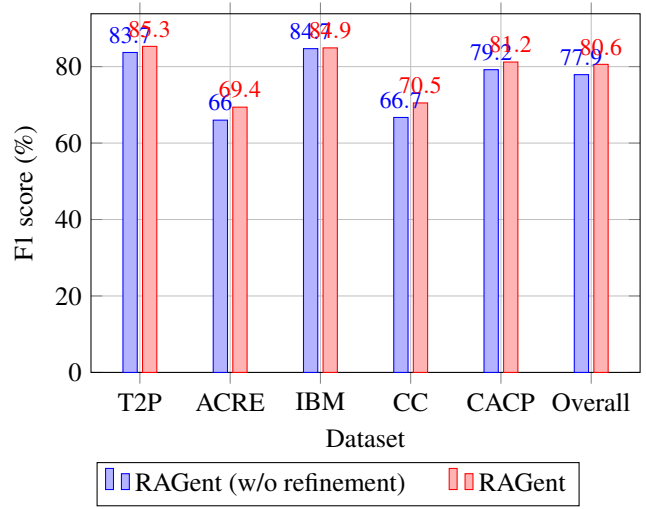

After the policies are generated from Step 4 and verified in Step 5, they are refined based on the verification result automatically, as explained in Section \ref{sec:approach}.
The effect of this iterative refinement mechanism on the policy generation performance of \framework{} can be seen in Figure \ref{fig:ablation/refinement}.
According to Figure \ref{fig:ablation/refinement}, \framework{} achieved the highest ACR generation reliability improvement of 3.8\% for CC and the lowest improvement (0.2\%) for IBM through iterative refinement. 
Upon close inspection of the generations from \framework{}, we found that, in contrast to the other document folds, CC contains many ambiguous NLACPs, making it harder for \framework{} to generate policies accurately, avoiding hallucinations.
For instance, most NLACPs from CC do not contain required components like the subject (e.g., \emph{``In step 2, the full paper must be submitted by uploading it.''}).
Even if there is a subject, it is often the website name \emph{``CyberChair''} without mentioning the exact entity like \emph{``site manager''}, compared to the subjects like \emph{``doctor'', ``professor'', ``nurse''} from other document folds.
Consequently, \framework{} has to depend on the verification and iterative refinement process to generate the correct policy frequently, leading to a significant accuracy drop if \framework{} does not refine the policies.

Figure \ref{fig:ablation/refinement} further shows that the iterative refinement also improves the \framework's reliability by $\approx 3\%$ when it comes to the \emph{``Overall''} dataset.
The above findings suggest that the reliability of the access control policy generation can further be improved through the \framework's novel automatic iterative refinement mechanism, as it reduces the ``hallucinations'' of deep learning models like LLMs \cite{kambhampati2024llms, skreta2023errors}.

\end{document}